\documentclass[12pt,preprint]{aastex}

\usepackage{natbib}
\usepackage{amsmath}
\shorttitle{MID-IR TO X-RAY OBSERVATIONS OF A STRONG SGR A* FLARE}
\shortauthors{DODDS-EDEN ET AL.}

\begin{document}

\title{Evidence for X-ray synchrotron emission from simultaneous mid-IR to X-ray observations of a strong Sgr A* flare}
\author{K.~Dodds-Eden\altaffilmark{1,*}, D.~Porquet\altaffilmark{2}, G.~Trap\altaffilmark{3,4}, E.~Quataert\altaffilmark{5}, X.~Haubois\altaffilmark{6},  
S.~Gillessen\altaffilmark{1}, N.~Grosso\altaffilmark{2}, E.~Pantin\altaffilmark{3,7}, H.~Falcke\altaffilmark{8,9}, D.~Rouan\altaffilmark{6}, 
R.~Genzel\altaffilmark{1,10}, G.~Hasinger\altaffilmark{1}, A.~Goldwurm\altaffilmark{3,4}, F.~Yusef-Zadeh\altaffilmark{11},  Y.~Clenet\altaffilmark{6}, 
S.~Trippe\altaffilmark{1}, P.-O.~Lagage\altaffilmark{3,7}, H.~Bartko\altaffilmark{1}, F.~Eisenhauer\altaffilmark{1}, T.~Ott\altaffilmark{1}, T.~Paumard\altaffilmark{6}, G.~Perrin\altaffilmark{6}, F.~Yuan\altaffilmark{12}, T.K.~Fritz\altaffilmark{1}, L.~Mascetti\altaffilmark{1} }

\altaffiltext{*}{katie@mpe.mpg.de}
\altaffiltext{1}{Max Planck Institut f{\"u}r Extraterrestrische Physik, Postfach 1312, D-85741, Garching, Germany.}
\altaffiltext{2}{Observatoire astronomique de Strasbourg, Universit{\'e} de Strasbourg, CNRS, INSU, 11 rue de l'Universit{\'e}, F-67000 Strasbourg, France}
\altaffiltext{3}{CEA, IRFU, Service d'Astrophysique, Centre de Saclay, F-91191 Gif-surYvette, France.}
\altaffiltext{4}{AstroParticule et Cosmologie (APC), 10 rue Alice Domont et Leonie Duquet, F-75205 Paris, France.}
\altaffiltext{5}{Department of Astronomy, University of California, Berkeley, 601 Campbell Hall, Berkeley, CA 94720-3411.}
\altaffiltext{6}{LESIA, Observatoire de Paris, CNRS, UPMC, Universit{\'e} Paris Diderot; 5 Place Jules Janssen, 92190 Meudon, France}
\altaffiltext{7}{Laboratoire AIM, CEA - Centre de Saclay, F-91191 Gif-sur-Yvette, France.}
\altaffiltext{8}{Department of Astrophysics, Institute for Mathematics, Astrophysics and Particle Physics, Radboud University, P.O. Box 9010, 6500 GL Nijmegen, The Netherlands}
\altaffiltext{9}{ASTRON, Oude Hoogeveensedijk 4, 7991 PD Dwingeloo, The Netherlands}
\altaffiltext{10}{Department of Physics, University of California, Berkeley, 366 Le Comte Hall, Berkeley, CA 94720-7300}
\altaffiltext{11}{Department of Physics and Astronomy, Northwestern University, Evanston, Il. 60208}
\altaffiltext{12}{Shanghai Astronomical Observatory, Shanghai 200030, China}

\keywords{accretion, accretion disks --- black hole physics --- infrared: general --- radiation mechanisms: general --- Galaxy: center --- X-rays: general}

\begin{abstract}
This paper reports measurements of Sgr A* made with NACO in L$'$-band (3.80 $\mu$m), Ks-band (2.12 $\mu$m) and H-band (1.66 $\mu$m) and with VISIR in N-band (11.88 $\mu$m) at the ESO VLT\footnote{The Very Large Telescope (VLT) at the European Southern Observatory (ESO) on Paranal, Chile: Program IDs 179.B-0261(A) and Program ID: 079.B-0929(A).}, as well as with XMM-Newton at X-ray (2-10 keV) wavelengths. On 4 April, 2007, a very bright flare was observed from Sgr A* simultaneously at L$'$-band and X-ray wavelengths. No emission was detected using VISIR. The resulting SED has a blue slope ($\beta > 0$ for $\nu L_\nu \propto \nu^{\beta}$, consistent with $\nu L_\nu \propto \nu^{0.4}$) between 12 micron and 3.8 micron. 

For the first time our high quality data allow a detailed comparison of infrared and X-ray light curves with a resolution of a few minutes. The IR and X-ray flares are simultaneous to within 3 minutes. However the IR flare lasts significantly longer than the X-ray flare (both before and after the X-ray peak) and prominent substructures in the 3.8 micron light curve are clearly not seen in the X-ray data. From the shortest timescale variations in the L$'$-band lightcurve we find that the flaring region must be no more than 1.2 $R_S$ in size. 

The high X-ray to infrared flux ratio, blue $\nu L_\nu$ slope MIR to L$'$-band, and the soft $\nu L_\nu$ spectral index of the X-ray flare together place strong constraints on possible flare emission mechanisms. We find that it is quantitatively difficult to explain this bright X-ray flare with inverse Compton processes. A synchrotron emission scenario from an electron distribution with a cooling break is a more viable scenario.

\end{abstract}

\maketitle

\section{Introduction}
The radio source coincident with the gravitational center of the Milky Way, named Sgr A*, was first discovered by Balick \& Brown in \citeyear{Balick&Brown_1974}.  It had already been suggested \citep{LyndenBell&Rees_1971} that the Milky Way may host a supermassive black hole at its center, and the newly discovered, unresolved source looked like it could well be the manifestation of such an object. That there really is a supermassive black hole of $\sim 4 \times 10^{6} M_\odot$, has now been proven beyond reasonable doubt through long-term monitoring and observation of the cluster of stars orbiting within arcseconds of the black hole, most notably the star S2 \citep[S0-2 in][]{Ghez_etal_2003} which has completed a complete 15-year orbit since the first monitoring observations in 1992 \citep{Schoedel_etal_2002_Nature, Ghez_etal_2003, Eisenhauer_etal_2005}.   

Sgr A* is thus a source of intense observational and theoretical interest, since it provides an avenue by which to study the  physics of accretion in the presence of extreme gravitational fields. However, it is unusually dim for a supermassive black hole \citep{Rieke&Lebofsky_1982}.  The spectral energy distribution of the radio source rises from radio towards submm wavelengths, but no steady emission can be detected above roughly $10^{12}$ Hz, implying that the spectral energy distribution (SED) turns abruptly around at this point (this feature has been named the `submm bump'). The overall luminosity is far below (by a huge factor of $\sim 10^{8}$) that expected for a black hole accreting at the Eddington rate. 

It was only recently that the source was discovered at all on the high frequency side of the submm bump, where it was found to exhibit strong flares in the X-ray \citep{Baganoff_etal_2001_FirstXrayFlare} and in the near-infrared (NIR) \citep{Genzel_etal2003_FirstNIRFlare}. A steady quiescent state in the X-rays at very low luminosities was also found  \citep{Baganoff_etal_2003}. The quiescent state has never been detected unambiguously in the NIR, nor has it ever been detected at mid-infrared (MIR) wavelengths for which only upper limits can be determined on either the quiescent state or possible flaring activity \citep[see for example, ][]{Schoedel_etal_2007}. 

 Subsequent to the first detections of Sgr A* flaring in the X-ray and NIR, a number of flares have been observed in both IR and X-ray wavelengths. Multiwavelength campaigns co-ordinating telescopes across the electromagnetic spectrum have worked towards obtaining simultaneous observations. 

Some general properties concerning the IR and X-ray flares that have emerged from those studies are:
\begin{enumerate}
 \item IR/NIR flares occur on average $\sim 4$ times per day \citep[see e.g. Figure 18,][]{Eckart_etal_2006_SubmmIRXray}, or between 30-40\% of the observing time \citep{YusefZadeh_etal_2006}.
 \item Strong X-ray flares occur on average $\sim~1$ per day \citep{Baganoff2003HEAD}. However, an enhanced rate of X-ray flaring can be observed within a time interval of roughly half a day \citep[e.g. a bright flare followed by three flares of more moderate amplitude][]{Porquet_etal_2008}. 
 \item Every X-ray flare appears to be associated with a NIR flare, however not
every NIR flare is associated with an X-ray flare \citep[e.g.,][]{Hornstein_etal_2007}. 
 \item X-ray and NIR flares occur simultaneously, with no significant delay \citep{Eckart_etal_2004,Eckart_etal2006_Multiwavelength,YusefZadeh_etal_2006}.
 \item Substructural variations with characteristic timescales of 15-25 minutes are seen in IR flares on a regular basis \citep{Genzel_etal2003_FirstNIRFlare,Meyer_etal_2006, Eckart_etal2006_Multiwavelength, Trippe_etal2007_PolarizedFlare}.
 \item Significant drops in flux are sometimes seen during X-ray flares \citep{Baganoff_etal_2003,Porquet_etal_2003}.
 \item Polarimetric investigations of the flares in the NIR have shown that the source
 is significantly polarized \citep{Eckart_etal2006_PolFlares} and that the polarization angle can swing in the tail end of the flare \citep{Trippe_etal2007_PolarizedFlare, Meyer_etal_2006}. 
\item At high fluxes the flare has a constant blue spectral index in $\nu L_\nu$\footnote{Here and elsewhere in this paper we use $\beta$ to denote the $\nu L_\nu$ spectral index, defined as $\nu L_\nu \propto \nu^{\beta}$.} of $\beta = 0.4$ between 3.8 and 1.6 $\mu$m \citep{Hornstein_etal_2007, Gillessen_etal_2006_VariationsInSpectralSlope}. For low fluxes it appears that the flare shows red $\nu L_\nu$ spectral indices \citep{Ghez_etal_2005_SgrAColor, Eisenhauer_etal_2005, Krabbe_etal_2006} with a possible trend of spectral index with flux \citep{Gillessen_etal_2006_VariationsInSpectralSlope} although this is disputed \citep{Hornstein_etal_2007}. 
 \item The two brightest X-ray flares \citep{Porquet_etal_2003,Porquet_etal_2008} have been observed to have well constrained soft $\nu L_\nu$ spectral index values $\beta = 0.2\pm0.3$ and $\beta=0.3\pm 0.3$ calculated at a 90\% confidence range \citep{Porquet_etal_2008}. While several fainter flares were observed, only a small number of photon index values has been reported; the latter exhibiting harder spectral indices \citep[e.g., ][]{Baganoff_etal_2001_FirstXrayFlare}. A re-analysis of XMM-Newton archival flares performed by \citet{Porquet_etal_2008} with a homogeneous data analysis shows that at low X-ray flux the spectral index is in fact not well constrained and a soft index as found for the two brightest flares cannot be excluded. Similarly, Mascetti et al. 2008 (submitted) analyses a co-added spectrum of all Chandra flares to date and reaches the same conclusion (i.e. that soft $\nu L_\nu$ indices are not excluded). Therefore, higher S/N spectra for individual weak/moderate X-ray flares are still required to establish whether all flares have similar spectral shape or not.
\item  The X-ray flares appear unambiguously to be `events', i.e. short, large amplitude outbursts followed by what looks like a perfectly flat baseline \citep{Baganoff_etal_2003, Porquet_etal_2008}. In the infrared, it is less clear whether this picture applies or whether the IR `flares' are simply peaks within an underlying sea of variability with the characteristics of red noise. Similarly, it is debated whether the substructural features seen in IR flares correspond to a characteristic frequency of the system (a quasi periodic oscillation (QPO)), or whether it is caused by statistical fluctuations in a smooth, red noise power spectrum \citep{Do2009,Meyer2008}.
\end{enumerate}

Note that there are also many important results from observations at longer wavelengths, but since we are directly concerned with the IR and X-ray data we have obtained, we have not gone into them in this paper. The full results of our April 2007 multiwavelength campaign including the observations at radio and submm wavelengths will be presented in Yusef-Zadeh et al. (2009, in prep.).

The quiescent state of Sgr A* can be successfully described by either a radiatively inefficient accretion flow \citep[or RIAF; see for example][]{Yuan_etal_2003}, or as arising from the base of a compact jet \citep{Falcke&Markoff_2000}. Each of these models can describe the observed properties of the quiescent state with similar magnetic field strengths ($B \approx 30$ G) and electron energies ($\gamma = E/mc^2 \approx 10$). The addition of Bremstrahlung emission from within the Bondi accretion radius ($R_{\text{Bondi}}\sim 1\arcsec$) explains the X-ray quiescent emission \citep{Quataert2002_Bremstrahlung}.

The origin of the flare emission within either of these basic pictures is much less certain. The high degree of linear polarization of the flares at IR wavelengths points to a synchrotron origin, but the emission mechanism responsible for the X-ray flares is not known. 
In the analysis of the simultaneous IR/X-ray multiwavelength observations to date, inverse Compton scattering processes have been favored. \citet{Eckart_etal_2004} and \citet{Eckart_etal2006_Multiwavelength} explained the simultaneity and the observed fluxes of infrared and X-ray flares through the synchrotron self Compton (SSC) emission of a compact source component emitting primarily at mm/submm wavelengths, with the emission at IR wavelengths possibly due to a combination of synchrotron and SSC emission. In \citet{YusefZadeh_etal_2006}, the IR and X-ray observations were interpreted within a picture where the X-ray emission was due to inverse Compton scattering of submm and IR photons involving populations of both submm-emitting and IR-emitting electrons. \citet{LiuMeliaPetrosian_2006} and \citet{Yuan_etal_2003} also present models involving inverse Compton processes. Synchrotron models for the X-ray flare have been suggested by \citet{Markoff_etal_2001} and \citet{Yuan_etal_2003} and IR/X-ray synchrotron models by \citet{Yuan_etal_2004}. Synchrotron models in general have been criticized due to the fact that the high energy electrons needed to generate X-ray synchrotron emission have very short cooling timescales (much shorter than the typical X-ray flare duration), requiring continuous injection in order to replenish the high energy population. However, this may not be such a disadvantage, and continuous injection is in fact a natural and reasonable expectation for the kinds of processes responsible for particle acceleration such as magnetic reconnection, turbulence and shocks. 

Going beyond the emission process behind the flare, there are models which attempt to simultaneously describe the detailed properties at one wavelength, such as the hot spot model \citep{BroderickLoeb_2005, Meyer_etal_2006_hotspotmodel, Trippe_etal2007_PolarizedFlare, Hamaus2009}, or accretion instability models \citep{Tagger&Melia_2006, Falanga_etal_2008} 

In this paper we present our multiwavelength observations and focus on constraining the emission mechanisms responsible for the simultaneous IR and X-ray flares we have observed. Although a full analysis of the detailed time-resolved SED evolution is beyond the scope of this paper, our high quality, full coverage, lightcurves in both L$'$-band and X-ray wavelengths offer the for the first time the opportunity to undertake detailed modelling of the time evolution of the flare SED, which may shed further light on the emission mechanisms and physical conditions/processes giving rise to a flare event.

Throughout this paper we adopt a Galactic Center distance of 8 kpc \citep{Eisenhauer_etal_2003}, and a black hole mass of $4\times10^6$~M$_{\odot}$ \citep{Gillessen2009,Ghez_etal_2008} for which the Schwarzschild radius is $R_S=1.2\times10^{12}$~cm.
 For the solar luminosity we used the value $L_\odot=3.8 \times 10^{33}$ erg s$^{-1}$.

\section{Observations}
In this section we present IR/NIR (3.8, 2.1 and 1.6 $\mu$m), MIR (11.88 $\mu$m) and X-ray (2-10 keV) observations of Sgr A* taken in April 2007. In particular we focus on April 4, 2007, on which date a very bright flare was observed in both L$'$-band (3.8 $\mu$m) and X-ray simultaneous to the MIR observations.
\subsection{IR/NIR Observations}\label{NIR_Observations}

The IR/NIR observations were taken at
the VLT in Chile as part of a multiwavelength campaign (LP 179.B-0261) in April
2007, using the NAOS-CONICA instrument
\citep{Lenzen_etal2003_NACO, Rousset_etal_2003_NACO} in imaging and polarimetric modes. We observed between 5:00 and 11:00 UT on April 1 to April 6 obtaining data in L$'$ (3.8 $\mu$m), Ks (2.1 $\mu$m) and H (1.6 $\mu$m) wavelength bands.

We subjected the raw data to a sky subtraction computed from
jittered object images in the L$'$-band case, and from dedicated observations of a patch of sky devoid of stars $\approx 700\arcsec$ W and $400\arcsec$ N of the GC for the Ks and H band observations. This was followed by flat-fielding and a correction
for dead/hot pixels. 

Once we had reduced the set of images, the raw
flux at the position of Sgr A*\footnote{confused in these observations with the star S17.} in each image was determined via two independent methods:
(i) aperture photometry, where the flux was computed as the sum of all
pixels within a small aperture centered on Sgr A*, from which the sum of pixels (normalized by area) within a larger annular region surrounding Sgr A* was
subtracted to remove background contamination; and
(ii) PSF photometry, where we used
StarFinder \citep{Diolaiti_etal2000_StarFinder} to automatically
identify and extract PSFs from the reduced images, thereby obtaining
source fluxes.

Finally, we calibrated the raw flux with the fluxes of
nearby stars of known and stable brightness, and converted it to a physical flux. For the extinction correction we used the values $A_L = 1.8$, $A_K=2.8$ and $A_H=4.3$ \citep{Genzel_etal2003_FirstNIRFlare}.

For those observations taken in polarimetric mode, we added the fluxes obtained in ordinary and extraordinary images to obtain an integrated flux \citep[for further details see][]{Trippe_etal2007_PolarizedFlare}.

\begin{figure}
\begin{center}
\includegraphics[width=0.5\columnwidth]{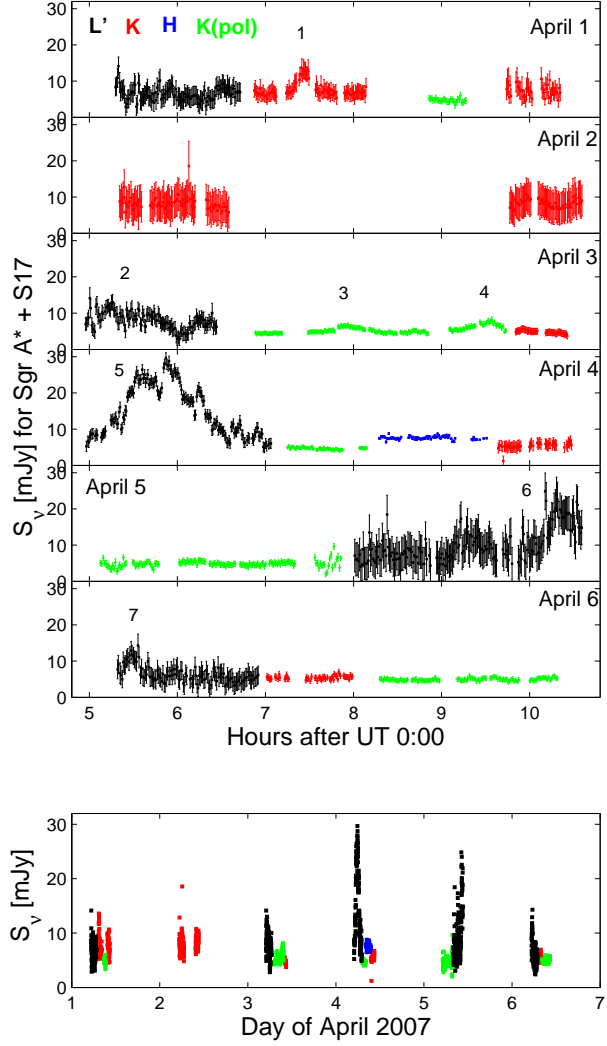}
\caption{Lightcurve for Sgr A* + S17 for April 1-6 2007 IR/NIR observations with NACO at the VLT. Observations were taken at L$'$ (black), Ks (red), H (blue) as well as in Ks-band using polarimetric mode (green). Several flare events are seen over the six nights of observations, labelled 1-7. Some nights show a more continuous level of variability (April 3), while on other nights there are long periods with no obvious variable emission. The L$'$-band flare from April 4 is the most significant event seen. A flare of equivalent strength in Ks band (given a colour of $\beta=0.4$ in $\nu L_\nu$) would reach $S_K \sim 20$ mJy. On April 5, another L$'$-band flare is seen under less favorable conditions which reaches $S_{L'}\sim 20$ mJy in L$'$-band.} \label{allApril}
\end{center}
\end{figure}

\begin{figure}
\includegraphics[width=1\columnwidth]{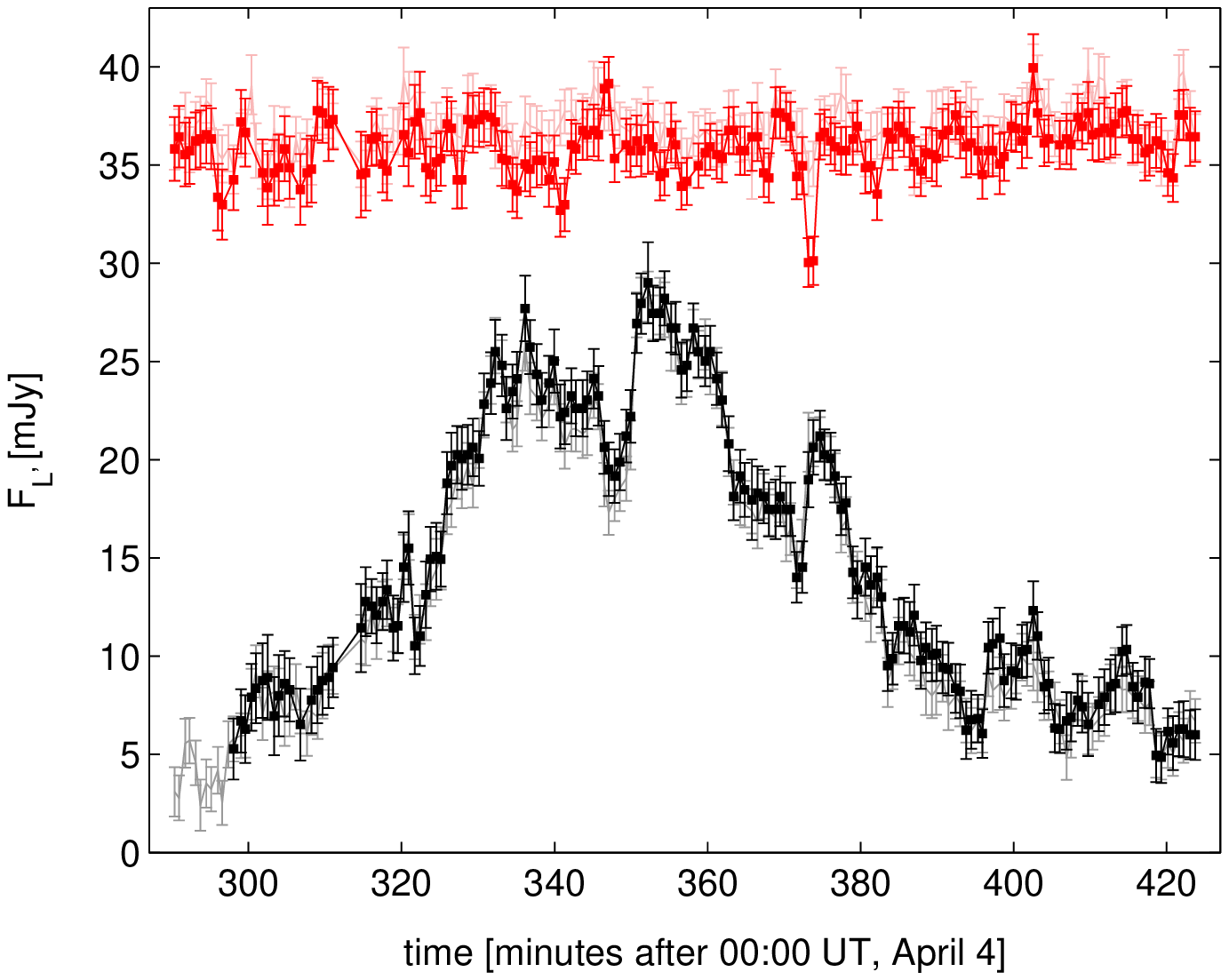}
\caption{Extinction-corrected L$'$-band flux of Sgr A* on 4 April 2007, determined via PSF photometry (black data points) and via aperture photometry (grey data points). Above the lightcurve of Sgr A* is shown in red (PSF photometry) and light red (aperture photometry) the flux of the nearby star S2 as a flux comparison, clearly demonstrating that the substructure is intrinsic to the Sgr A* source. S2 was confused at the time of observation
with S13, and in the figure their combined flux is shifted upwards by 25 mJy. Sgr A* was confused with S17 and the flux shown also possibly has a contribution from a dust cloud as well as the unknown quiescent state of Sgr A*. The data are binned to a bin-width of 44.3s. The PSF photometry method only lists fluxes of sources detected with 3$\sigma$ significance, which explains why the lightcurve derived by PSF photometry begins only at around $\sim$ 300 minutes. We use the mean of these points with $t \lesssim 300$ minutes as an upper estimate of the background level (S17 + confused sources + quiescent state) upon which the flare emission is superimposed. If the minimum of the lightcurve is used instead, we obtain a background estimate of 2.4 mJy.} \label{IRflare}
\end{figure}

The resulting lightcurves for the combined source Sgr A* + S17 are presented in Figure \ref{allApril}.
Several weak flares are seen (labelled 1-4 and 6-7).
On the night of 4 April and under good conditions (seeing $\sim
0.55-0.9$ and Strehl ratios $\sim 0.45-0.65$), a very strong flare
was seen in L$'$-band at the position of Sgr A*,
beginning just before 06:00 UT, April 4, and lasting for roughly 2
hours. We present the lightcurve of this flare in detail in Figure \ref{IRflare}.

Since Sgr A* was confused with S17 on April 4 2007, S17 also contributes flux to the lightcurve shown in Figure \ref{IRflare}. In addition, the quiescent state of Sgr A*, if it exists, is not well known so it is possible that the quiescent state and possibly other L$'$-band sources such as a small dust cloud close to Sgr A* \citep{Clenet_etal_2005} also contribute to the minimum flux of Sgr A* on the night of April
4. It is also not clear from the lightcurve whether Sgr A* ever reaches a level of non-activity during our L$'$-band observations. If we take the mean of the group of points where the lowest flux for the night was recorded between $t \sim 295$ to $300$ minutes, this leads to an upper estimate for the contribution of any nonvariable emission of $4.3$~mJy.

The combination of the data quality and the strength of the flare
activity make the April 4 flare presented
here the best specimen of our entire dataset recording the L$'$-band
activity of Sgr A* and spanning 2003 to 2007. The lightcurve shows
very significant substructure on a timescale of $\sim$ 20 minutes. This kind of substructure has been seen in previous Ks-band
flares
\citep{Genzel_etal2003_FirstNIRFlare,Eckart_etal2006_PolFlares,
Trippe_etal2007_PolarizedFlare}, but is seen here for
the first time in L$'$-band. This strengthens the case that the
presence of substructure is a common feature of IR/NIR flares.

\subsubsection{Limits on L$'$-band flaring activity $\sim$~7-11~hrs~UT.}
\label{Section_LimitsOnFlareActivity}
On April 4 2007 we also obtained data in other wavelength bands (Ks-band polarimetry, H-band imaging and Ks-band imaging) for another $\sim$ 3.5 hours after the flare observed in L$'$-band (Figure \ref{allApril}). There was no obvious variability in these wavelength bands. We find that our measured Ks-band flux for Sgr A* + S17 of $S_K = 5.1 \pm 0.4$ mJy is consistent with the measurement of Do et al (2008) for S17 alone (5.5 mJy, dereddened according to $m_K = 2.8$ as assumed in this paper; no error was given). It is also consistent with our own past measurements of S17's magnitude ($m_K=4.8$ mJy), although our error on this value is very large and on the order of $2$ mJy. The H-K color of the combined Sgr A* + S17 source is $\beta = 2.6\pm0.7$, consistent with that of a pure stellar source ($\beta=3.0$). From this and the lack of significant variability in the emission we would conclude that in H and Ks band the emission is dominated by S17. We note that in deconvolved images of this dataset we \emph{do} see an elongation of the source in Ks band, and resolve two distinct sources in H band. Of these two H-band sources we can not be sure whether the source coincident with Sgr A* is stellar (e.g. from faint unresolved S-stars surrounding the black hole) or whether it might be quiescent/flaring emission. Due to the close proximity of the sources (only 3 pixel separation) it was not possible to determine the Ks-band fluxes of each source accurately.

In L$'$-band the lowest measured luminosity lies above the extrapolation of the (Sgr A* + S17) H-K measured color and we can conclude that within the timespan of our L$'$-band observations we did not reach the flux level of S17, which would be expected at $\sim 1$ mJy. The remaining flux we see may be due to a further contamination of the L$'$-band flux by a confused source (e.g. a small dust cloud known to be an L$'$-band source nearly coincident with Sgr A*, $\sim 4.7$ mJy Clenet et al 2005, although this seems unlikely given the high flux), or it might also be due to the fact that the flare activity never ceased within our L$'$-band observation time interval. Since flares are redder than the stars, we can not rule out that some low-level flaring continued to occur after $\sim 430$ minutes while we observed in Ks- and H-bands. 

To make some estimation of the L$'$-band flux during the time interval within which we observed in Ks and H bands we must extrapolate from our Ks-band measurements, which introduces large uncertainties. We can reasonably assume that any flaring emission was below $S_K \sim 2.1$ mJy during the Ks and H-band observations (our lower limit on the flux of S17 is 3 mJy; note this is also consistent with extrapolating the flux of the deconvolved, separated source in the H-band images to Ks-band with a slope of $\beta=3$). Although the flare color at low flux levels is not well established, if we take a flare color of $\nu L_\nu = 0.4$ we can estimate that the combined source of Sgr A* + S17 should not have been at a flux level higher than 4.7 mJy in L$'$-band during this time. However, if the flare were redder at low flux levels or if some flux is contributed from the dust cloud near Sgr A* then the limit on the flux level could be higher. We obtain an upper limit of 9 mJy if we add the lowest flux detected during the L$'$-band observations (4.3 mJy), using it as an upper limit on the magnitude of any nonvariable contamination. 

\subsection{X-ray Observations}

\begin{figure}
\begin{center}
\includegraphics[width=0.5\columnwidth]{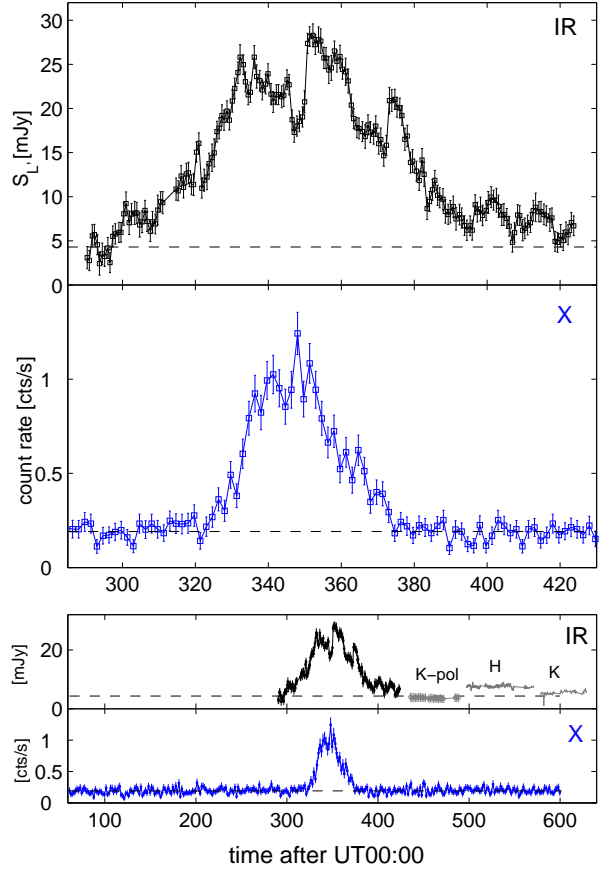}
\caption{Comparison of the L$'$-band and 2-10 keV
lightcurves (labelled IR and X, respectively). The top two panels show the two flares over the period of the  L$'$-band observations. In the lower two panels, a larger time interval is shown. We also show data taken in other wavelength bands (Ks-band polarimetry, H-band imaging, and Ks-band imaging) subsequent to the L$'$-band flare indicating that the flare activity ceased in both wavelength bands after roughly 4:00 UT.  In the same night of observations three more X-ray flares were seen (Porquet et al 2008); the first of these started at UT 11:32 (692 minutes), i.e. roughly an hour after the last of the NIR observations. The dashed lines indicate our estimates of the background levels (i.e. emission that is not flaring emission) at each wavelength. In the case of the IR lightcurve this may be an overestimate.} \label{comparison}
\end{center}
\end{figure}

On the 4th of April 2007, the VLT observations described in Section
\ref{NIR_Observations} overlapped with those of XMM-Newton. The
observations and data reduction of the X-ray flare are published by
Porquet et al (2008), and will not be repeated here. The X-ray
lightcurve is presented in Figure \ref{comparison}, where it is
compared with the L$'$-band lightcurve.

The X-ray flare was very bright. The 2-10 keV spectrum of the flare showed a soft spectrum: a power law fit correcting the underlying model for dust scattering and absorption gives a power law slope of $\Gamma = 2.3 \pm 0.3$ (error bars given at the 90$\%$ confidence level), equivalent to a $\nu L_\nu$ spectral index of $\beta = -0.3 \pm 0.3$. The bright flare observed on April 4th \citep[labelled \#2 in ][]{Porquet_etal_2008} is the second brightest flare observed so far from Sgr A* with an amplitude of about 100 compared to the quiescent state. \citet{Porquet_etal_2008} show that this flare and the brightest X-ray flare observed in October 2002 \citep{Porquet_etal_2003} have similar light curve shape, duration, and spectral characteristics (photon index).

\subsection{Mid-Infrared Observations}\label{MIR_Observations}
VISIR, the VLT Imager and Spectrometer for the mid-infrared, mounted
on the ESO/VLT telescope Melipal (UT3) at Paranal, Chile
\citep{Lagage_etal_2004_VISIR, Pantin_etal_2005}, observed the
Galactic Center from 2007-04-04 05:29:00 to 2007-04-04 10:34:00 (UT). We collected the data with
the imaging \texttt{PAH2\_2} filter on, at 11.88 $\pm$ 0.37 $\mu$m
in the atmospheric window ``N''. The Small Field mode (\texttt{SF})
was employed, resulting in a field of view of 256 $\times$ 256
pixels (19.2 arcsec$^2$), each pixel corresponding to 0.075
arcsec$^2$.

We performed the calibration of the \texttt{PAH\_2} filter on
2007-04-04 05:16:24 (UT) with a 109.9s observation of the standard
star HD 102461 \citep[9.237 Jy in the \texttt{PAH2\_2}
filter;][]{Cohen_etal_1999}.

The basic `chopping and nodding' technique was applied to acquire
the data, which were then reduced with the standard VISIR
pipeline\footnote{see
\texttt{http://www.eso.org/instruments/visir/}}: this involved
flatfielding, bad pixel correction and combination of a stack of
chopped and nodded frames to produce a final set of 79 consecutive
images.

We determined the position of Sgr A* using the precise positions of
the SiO maser sources IRS 7, IRS 9 and IRS 10EE, distributed about
Sgr A* \citep{Reid_etal_2007}. No point source at the position of
Sgr A* is detected in either the individual images or the collapsed
image of the entire night. We also performed a Lucy-Richardson
deconvolution with HD 102461 as PSF with again no source detection. The flux from a
box of 0.375 arcsec$^2$ centered on the position of Sgr A* is constant with an average value of 123 $\pm$ 6 mJy. This flux may be
attributed to the weak and diffuse dust ridge on which Sgr A* lies,
and our measured value is consistent with previous VISIR
observations
\citep{Eckart_etal_2004, Schoedel_etal_2007}.

To determine an upper limit of the brightness of the simultaneously
observed flare, we subtracted an average image of the
quiescent phase (12 images from 07:17 to 08:00) from an average
image of the flaring phase (12 images from 05:30 to
06:13\footnote{Note that VISIR observations started 5 min after the
beginning of the X-ray flare, which was from 05:25 to 06:13}). We
included a slight shift in the relative positions (less than half a
pixel) of the images, degraded the mean images with Moffat functions
to mimic the slight differences of atmospheric conditions between
them, and finally destriped the subtracted image. Over a region of
$\sim$ 3 arcsec$^2$ centered on Sgr A*, north of the minispiral, the
subtracted image displays a relatively flat background. 

To quantitatively estimate our detection limit, we proceeded by
simulating an artificial flare in the data. We included a weak point
source of a given flux (with VISIR's PSF) at the position of Sgr A* in the
substracted image. We increased the point source's flux until it was detected at a significance of 3$\sigma$ and took this value as an upper
limit on the flare's mean flux. We thus estimate that Sgr A* could not
have been brighter than $\sim$ 12 mJy at 11.88 $\mu$m (3$\sigma$, not dereddened). Note that this value is compatible
with VISIR's empirical sensitivity at this wavelength: 7
mJy/10$\sigma$/1 hr (median value for different atmospheric
conditions).

The value of the extinction correction in the MIR depends critically
on the strength and shape of the silicate absorption feature at
$\sim 9 \mu$m. The values in the literature are published as ratios
relative to $A_K$ or $A_V$, so we use the value $A_K = 2.8$ mag
($A_V=25$) mag to ensure consistency across our multiwavelength
observations. The closest extinction measurement to our observation
wavelength, $\lambda = 11.88 \mu$m, was made by
\citet{Lutz_etal_1999} for a wavelength of $\sim 12.4 \mu$m. We
consider three theoretical models
\citep{Chiar_etal_2006,Draine&Lee_1984,Roche&Aitken_1984} for the
shape of the silicate profile in the region, to allow us to
extrapolate the value measured at $12.4 \mu$m to $11.88 \mu$m. These
models each use different sources as template profiles but are all
very similar in slope around 12 $\mu$m and result in very similar
values of extinction when normalized to the \citet{Lutz_etal_1999}
value at 12.4 $\mu$m. The thus determined extinction value is
$A_{11.88 \mu\text{m}} = 1.7 \pm 0.2$ mag. With this value, the
dereddened 3$\sigma$ upper limit on emission from Sgr A* during the
flare is $F_\nu^{\text{MIR}} \sim 57$ mJy.

\section{Results}

\subsection{Simultaneity of infrared and X-ray flare}

We have obtained complete, fully sampled lightcurves in L$'$-band and X-ray. The X-ray lightcurve showed no other events during either a $\sim 15$ hour period before and $\sim$ 5 hours after the bright X-ray flare. In the IR/NIR, in the remaining observations of about 3$\frac{1}{2}$ hours following the L$'$-band flare, there were no other flares or obvious variable emission above a level of $S_K\approx2.4$ mJy (implying $<12\%$ the equivalent Ks-band flux of the peak of the L$'$-band flare, using $\beta=0.4$). We translated this to a more conservative limit (see Section \ref{Section_LimitsOnFlareActivity}) in L$'$-band of $S_{L'}<9$ mJy ($<32\%$ the peak L$'$-band flare flux).  At least on this occasion, both infrared and X-ray emission can thus be best characterized as isolated `flare' events.

From a correlation analysis, the L$'$-band and X-ray flares are found to be simultaneous to within $\sim 3$ minutes. In particular we do not see any significant delay or asymmetry in the longer wavelength emission relative to the peak of the X-ray flare, thus excluding that adiabatic expansion of an initially optically thick blob plays a role in the infrared and X-rays.

\subsection{General Lightcurve Shape}
This multiwavelength observation allows us to make the most detailed lightcurve comparison so far, of simultaneous IR and X-ray flares from Sgr A*. From the comparison of the two lightcurves shown in Figure \ref{comparison} it appears that the L$'$-band flare begins first. The L$'$-band lightcurve rises before any significant X-ray emission is seen, and L$'$-band emission remains after the X-ray emission has subsided. It appears that the two events have different durations from one another, and that the infrared event lasts longer overall than the X-ray event.

Taking the uncertainties on the background levels into account, we measure FWHM durations for each (background-subtracted) lightcurve of FWHM$_{IR} = 66 \pm 8$ and FWHM$_X = 28 \pm 0.5$ minutes. Thus we find that the FWHM of the L-band flare is $\sim 2$ times that of the X-ray flare. 

\subsection{Substructure} \label{substructure}

One very striking feature of the L$'$-band lightcurve is the substructural variations seen on a timescale of $\sim 20$ minutes. The variations in flux are large: up to $\sim$ 30\% the peak flux. There are no apparent features that would correspond to these in the simultaneous X-ray lightcurve.

Although the X-ray lightcurve has lower SNR than the L$'$-band lightcurve, the error bars are only on the order of $\sim$ 10\% the peak flux, and thus cannot hide substructures as large as those in the L$'$-band lightcurve. 
From this we conclude that the lack of substructure in the X-ray lightcurve is not due to lower SNR, and that this property is in fact intrinsic to the simultaneous IR/X-ray lightcurves.

\subsection{Shortest time-scale variations}\label{Section_ShortTimeScaleVariations}

In the L$'$-band lightcurve, in particular at $t \sim 350$ minutes but also near to  $t\sim 370$ and $395$ minutes, very rapid changes in flux (factors $120\%$ to $170\%$, significance $> 3\sigma$) are observed within a very short timescale, $\Delta t < 47 $ seconds.

Such short term variations place a limit on the size of the flaring source, or at least the size of the part of the source providing the sudden change in flux (which is a significant fraction, $\sim 30\%$, of the total flux). Since such variations cannot propagate within the source faster than the speed of light $c$, the source size $R_F$ is immediately constrained to be

$$R_F < c \Delta t  = 1.2 R_S.$$

A caveat to this constraint is that we have not considered various relativistic factors; given the small size obtained and the fact that we think the flare might occur at very small radii in the accretion flow, relativistic effects might be an important factor in influencing the time variability of the source. Relativistic beaming near the event horizon is a possible source of magnifying the amplitude of variations due to an underlying spatial structure in the infrared emission. Such beaming effects could be considerable \citep{Hamaus2009,BroderickLoeb_2005}.

\subsection{Power Spectra}

Whether or not the substructures seen in the L$'$-band lightcurve are indicative of a QPO or are merely spurious peaks in a red noise process is a matter of current debate \citep[e.g.][]{Do2009,Meyer2008}. Since the putative QPOs inevitably turn out to be too weak to stand a significance test from a single observation night's worth of data, we turn our focus to longer timescales. H-band and Ks-band polarimetric data that were taken following the L$'$-band measurements show no evidence of any variable emission, and this clearly holds an implication for the variability behaviour of the source on longer timescales. 

\begin{figure}
\includegraphics[width=\columnwidth]{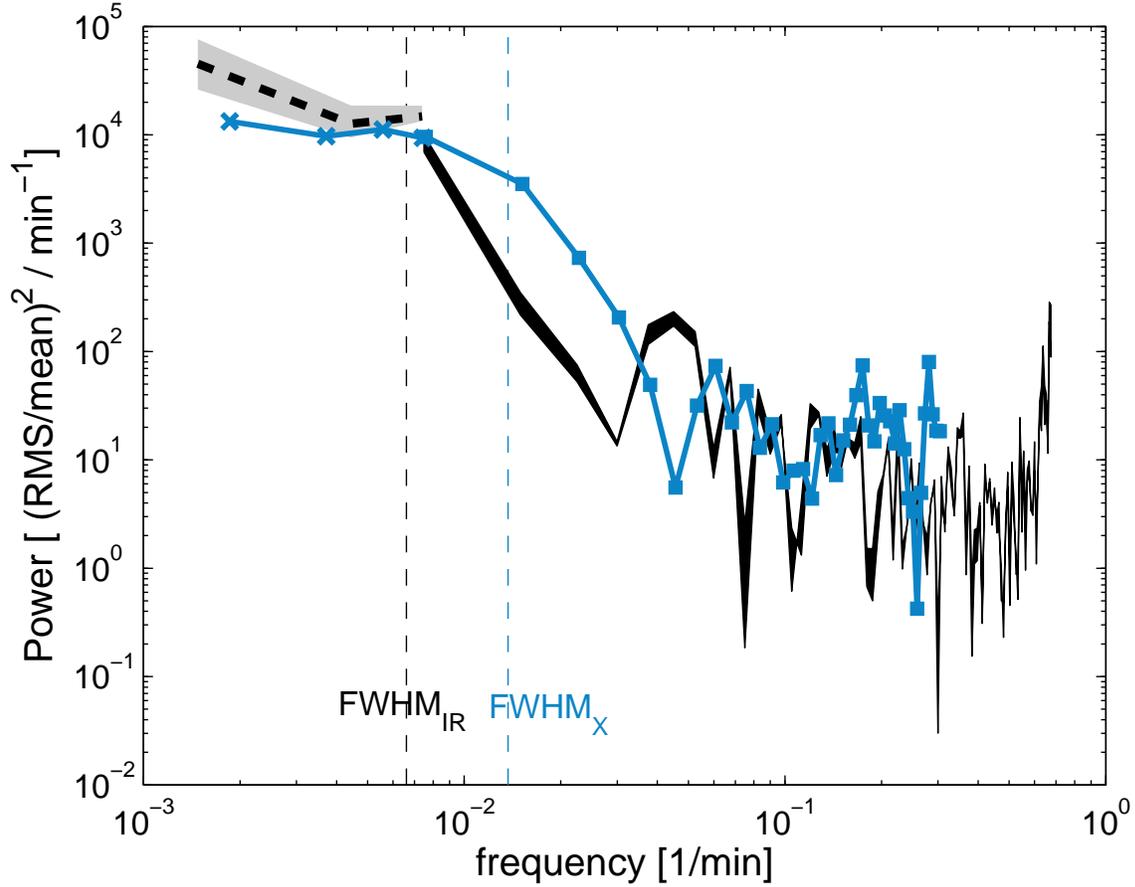}
\caption{Power spectra of the L$'$-band (thick black) and X-ray (blue squares) lightcurves. Our constraint on the periodogram at low frequencies from Ks and H band data is shown as the gray region (with the interval mean shown as black dashed line). The RMS-squared normalization was used, where the mean of the IR and X-ray lightcurves was taken from the same time interval. A peak is seen around the 20 min timescale ($\sim 0.04$ min$^{-1}$), while there is no corresponding peak in the X-ray power spectrum.
The two dashed lines indicate the corresponding FWHM frequency for the Fourier transform of a Gaussian given the FWHM durations of the IR and X-ray flares.
}
\label{powerspectra}
\end{figure}

Figure \ref{powerspectra} shows the periodogram of the L$'$-band data compared with that of the X-ray lightcurve. We use the ordinary periodogram with the RMS-squared normalization \citep[see, for example][]{Uttley2002} which allows us to compare lightcurves taken with different instruments (and here at different wavelengths). For a consistent comparison between the IR and X-ray variability we took the mean from the same time interval, i.e. the maximum time overlapping time interval of the IR and X-ray observations. For timescales $\lesssim$ 130 minutes (frequencies $>$ 0.008 min$^{-1}$) we show the power spectrum of the L$'$-band data only. We use our limits on the variable emission in Ks and H bands to constrain the periodogram at lower frequencies. Some uncertainty in the normalization of the IR power spectrum comes about through our uncertainty in the mean value given the extrapolation from Ks-band to L$'$-band. There is an apparent peak at $\sim$ 20 minute timescales. Whether it is a real QPO or the spurious peak of a red noise spectrum, it is noteworthy that the putative QPO peak of the L$'$-band data has no corresponding feature in the X-ray power spectrum. This is consistent with our observation that the substructures of the IR lightcurve are not present in the X-ray lightcurve, which is comparatively smooth. 

 At low frequency we also see the difference in widths of our lightcurves; the power spectrum of the X-ray lightcurve resembles a Gaussian at low frequencies which is as expected for the power spectrum of a single Gaussian-like flare event. The power spectrum of the IR lightcurve resembles a narrower Gaussian, again expected from the fact that the IR lightcurve was of longer duration than the X-ray lightcurve. We note then that the clear flattening of the power spectrum towards low frequencies again suggests that the IR flares are discrete events. 

\subsection{Spectral Energy Distribution}
\begin{figure*}
\begin{center}
\includegraphics[scale=0.5, angle=-90]{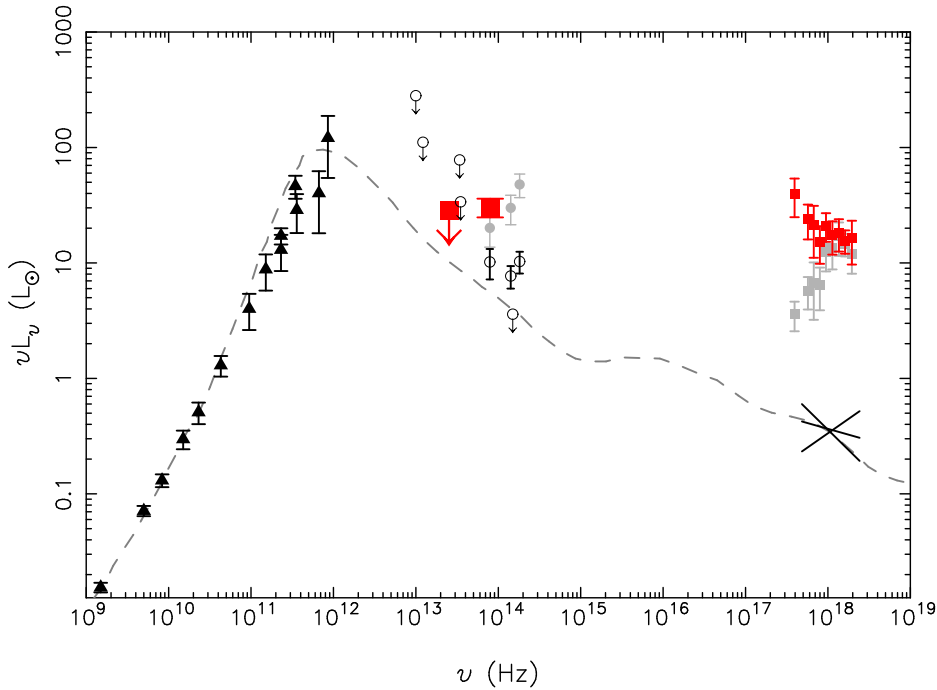}
\caption{The Spectral Energy Distribution of Sgr A*: in black (filled triangles) are radio to submm measurements of the quiescent state \citep{Markoff_etal_2001, Zhao_etal_2003}. Note that these measurements are time averaged measurements and the errorbars include variable emission of up to 50\%. As open black circles with arrows are shown 30 $\mu$m, 24.5$\mu$m and 8.6 $\mu$m upper limits taken from \citet{MeliaFalcke2001}, the upper limit at 8.6 $\mu$m from \citet{Schoedel_etal_2007} and the limit on the quiescent state at 2 $\mu$m from \citet{Hornstein2002}. The quiescent state values from \citet{Genzel_etal2003_FirstNIRFlare} are shown as the open black circles with errorbars, and the flare values from the same paper as gray filled circles.  The X-ray quiescent state is shown as the black bow-tie
\citep{Baganoff_etal_2003}. The dashed line shows a model for the quiescent state \citep{Yuan_etal_2003}. 
Our new measurements for the SED of a flaring state of Sgr A* are shown in red (filled squares): 
(i) the MIR (11.88 $\mu$m) upper limit is shown as the downwards-pointing arrow. The MIR upper limit is determined over an interval 05:30 to 06:13
(see discussion in Section \ref{MIR_Observations}). 
(ii) The L$'$-band measurement is shown as the red square with errorbars. This corresponds to extinction corrected, background subtracted mean value of the L$'$-band observations, $19.1 \pm 3.6$ mJy (in this case we used the minimum of the lightcurve, 2.4 mJy as the background estimate). Since the MIR observations did not start until 5:30 (approximately half an hour after the onset of the NIR flare), the mean was computed over the MIR time interval rather than the entire L$'$-band flaring interval. 
 Also shown next to the L$'$-band data point is how the L$'$-band measurement would continue into Ks and H band with a slope of 0.4, characteristic of the L$'$-H slope of a number of `bright' observed flares in the literature \citep{Hornstein_etal_2007,Gillessen_etal_2006_VariationsInSpectralSlope} and also consistent with the slope of the peak flare values of \citet{Genzel_etal2003_FirstNIRFlare}. (iii) Two possible X-ray spectra are shown, neither of which is model-independent. The red points which slope upwards to the left indicate the power law fit, of \citet{Porquet_etal_2008}, while the grey points show the blackbody fit of the same paper.  The X-ray spectrum was scaled by a small factor since it incorporated data from an extra five minutes before the MIR observations began.
}
\label{SED}
\end{center}
\end{figure*}

The `flare state' SED for the observations of Sgr A* on April 4, as determined by our multiwavelength observations, is
shown in Figure \ref{SED}.

For the $\nu L_\nu$  value at L$'$-band we computed the mean of the extinction corrected, background subtracted L$'$-band flux. We chose to take the mean value rather than the peak value since the MIR limit and X-ray spectra were both determined as averages over the flare interval. The error in the L$'$-band data point is computed as the standard deviation of the lightcurve. 

For the X-ray data, it is difficult to show an intrinsic, dust and absorption-corrected X-ray spectrum without the assumption of a model. This is because the inversion of a raw counts/channel X-ray spectrum is generally non-unique and unstable to small changes in the counts/channel spectrum \citep{Arnaud1996_XSPEC}. To determine the best fitting spectrum, a model is calculated and `folded', or convolved with the instrumental response after which the folded model spectrum is compared to the observed counts in each channel. Once one has found a best fit, the process can be reversed for the best fitting model and one obtains an intrinsic, but model dependent, spectrum.

 Because of this, we can not show a single, model-independent X-ray measurement on the SED for the April 4 flare. Instead, we show two possible X-ray spectra as obtained by \citet{Porquet_etal_2008} (i) assuming a power law shape, with $\Gamma = 2.3\pm 0.4$  and $N_H = 12.8_{-2.1}^{+2.5} \times 10^{22}$ cm$^{-2}$ \citep[at the 90$\%$ confidence level, using the $\chi^2$ statistic; see Appendix B in ][]{Porquet_etal_2008}, and (ii) assuming a blackbody model, with parameters $N_H=7.3_{-1.3}^{+1.6}\times 10^{22}$ cm$^{-2}$ and $kT = 1.5_{-0.1}^{+0.1}$ keV. The blackbody fit had the lowest $N_H$ of the models investigated in that paper. Both models show a soft spectral index above $\nu \sim 10^{18}$ Hz. We scaled the X-ray data, which was determined over the full X-ray flaring interval, by a factor 0.95 to account for the fact that the interval over which the MIR upper limit was calculated was shorter by 5 minutes (the scaling factor was determined as the ratio in fluxes between these two intervals).

In both L$'$-band and X-rays, there is a substantial increase in flux above the quiescent level. The absence of any detectable emission at 11.88 $\mu$m implies that the flare emission spectrum must rise from 11.88 $\mu$m to 3.80 $\mu$m. This appears
consistent with a ($\nu L_\nu$) spectral index of $\beta = 0.4$ \citep{Hornstein_etal_2007, Gillessen_etal_2006_VariationsInSpectralSlope, Genzel_etal2003_FirstNIRFlare}.

The rise in $\nu L_\nu$ from the MIR towards NIR wavelengths suggests that the population of electrons producing the L$'$-band flare must have a different distribution of electron energies to those in the submm bump. This might be an power law tail of transiently accelerated electrons, for example, or a small group of electrons heated to a high temperature. What this observation shows is that a NIR flare cannot be due to a small change in \emph{overall} properties of the submm bump (such as, for example, a global increase in magnetic field which temporarily increases the emitted synchrotron emission of the quiescent state). The flare event must involve only a small fraction of the quiescent state electrons, either in some kind of acceleration process that acts globally but inefficiently within the accretion flow, or via a more efficient but very localized acceleration process. As mentioned in Section \ref{Section_ShortTimeScaleVariations}, the short time scale variability of the source also points towards a localized event.

\section{Modelling the flare state SED of Sgr A*}
\label{Section_SEDmodelling}
We studied the flare state SED of Sgr A* under four simple SED models. These models explore different emission scenarios for the IR/X-ray flare and correspond to scenarios where the IR and X-ray flares are due to synchrotron and inverse Compton emission mechanisms. For the fitting of the SED models to the data we use the X-ray spectral fitting program XSPEC \citep{Arnaud1996_XSPEC}. To incorporate our infrared data points we add an extra data channel with the L$'$-band extinction corrected flux; the instrumental response for this data channel is an identity matrix. The four XSPEC models we used are:
\begin{enumerate}
 \item \texttt{icmodel}: IR emission is synchrotron emission by transiently heated/accelerated electrons; X-ray emission is due to the inverse Compton scattering of submm photons from the `quiescent' population of electrons by the population of electrons producing the IR emission.

 The seed photon spectrum comes from the quiescent population of electrons and its spectrum and total luminosity is fixed \citep[we use the model spectrum of][]{Yuan_etal_2003}. $R_Q$, the size of the region containing the quiescent state (submm-emitting electrons) electrons is a free parameter and controls the photon density of submm photons available for inverse Compton scattering. 

The IR synchrotron emission is modelled based on a thermal distribution of electrons. There are three parameters that pertain to the transiently heated population of electrons producing IR synchrotron emission: $B$, the magnetic field, $\theta_E$, the dimensionless electron temperature ($\theta_E$ denotes the typical energy $\gamma$ of the electron distribution; it is equal to $kT_e/mc^2$, where $T_e$ is the temperature of the thermal electron distribution), and $N$, the total number of IR synchrotron emitting electrons.

 \item \texttt{sscmodel}: IR emission is synchrotron emission by transiently heated/accelerated electrons; X-ray emission to IR/NIR photons of the transiently heated/accelerated (flare) electron population that are inverse Compton scattered by the same population (i.e. SSC).

The IR synchrotron emission is again modelled based on a thermal distribution of electrons. This IR synchrotron emission can be again computed from the parameters: $B$, magnetic field, $\theta_E$, the dimensionless electron temperature (the typical $\gamma$ of the electrons, see \texttt{icmodel}), and $N$, the total number of IR synchrotron emitting electrons. 

In this model it is the parameter $R_F$, the size of the region containing the flaring (IR-emitting) electrons that controls the photon density of the seed photon spectrum.

 \item \texttt{powerlaw}: IR emission is synchrotron emission from a power law energy distribution of accelerated electrons. The parameter of interest in this model is the particle index of the power law electron distribution, $p$, i.e. $N(\gamma) \propto \gamma^{-p}$.

 \item \texttt{powerlawcool}: IR emission is again synchrotron emission from an electron distribution with continuous injection of power law electrons and the addition of synchrotron cooling. Parameters of this model are $p$, the particle index of the injected electron spectrum, and the magnetic field $B$, which determines the cooling time of electrons and thus the energy/frequency at which the cooling break occurs.
\end{enumerate}
The two models \texttt{icmodel} and \texttt{sscmodel} were developed especially for this work, while \texttt{powerlawcool} was a simple adaptation of the existing XSPEC model \texttt{powerlaw} to incorporate the cooling break. Specific details of the \texttt{icmodel}, \texttt{sscmodel} and \texttt{powerlawcool} models are listed in the Appendix.

For each model we also take into account the effect of photoelectric absorption and dust scattering on the X-ray spectrum via the XSPEC routines \texttt{scatter} and \texttt{wabs} \citep[for more details, see][]{Porquet_etal_2008}. These effects were not applied to the NIR data. For the dust scattering (\texttt{scatter}), we fix $A_V=25$ to match the dust extinction corrections used for the L$'$-band and MIR data. For photoelectric absorption (\texttt{wabs}) we allow the parameter $N_H$ to be determined.

As an extra constraint, we add an extra data point at H-band (1.65 $\mu$m) which corresponds to a constraint on the $\nu L_\nu$ slope from L$'$-H band of $\beta=0.4\pm0.2$. We find that adding this constraint generally results also in models which do not violate the MIR 3$\sigma$ upper limit. We ran models also without this extra constraint and very similar best fit values were obtained.

\begin{deluxetable}{lllllllll} 
\rotate
\tabletypesize{\scriptsize}
\tablecolumns{9}
\tablewidth{0pc}
\tablecaption{Models: Fit Parameters\label{Table_FittedParameters}}
\tablehead{\colhead{} & \multicolumn{8}{c}{Fit to Mean Fluxes \& NIR spectral index} \\Parameter & \multicolumn{2}{c}{\texttt{icmodel}} &  \multicolumn{2}{c}{\texttt{sscmodel}} &  \multicolumn{2}{c}{\texttt{powerlaw}} &  \multicolumn{2}{c}{\texttt{powerlawcool}} }
\startdata
N$_H$ [$\times 10^{22}\ \textnormal{cm}^{-2}$]& $11.7$ & $(9.9,14.3)$ & $11.5$ & $(9.7,13.7)$ & $11.5$ & $(10.6,12.7)$ & $12.4$ & $(11.0,12.1)$ \\
B [Gauss]& $210$ & $(30,2900)$ & $6000$ & $(2200,7900)$ & \multicolumn{2}{l}{$<0.1$, or $>60$} & $6.1$ & $(0.1,60)$ \\
$\theta_E$ [$kT_e / m_ec^2$]& $140$ & $(50,210)$ & $11$ & $(9,16)$ & - &  & - & \\
$N_e$ [$\times 10^{40}$ electrons]& $4.7$ & ($0.2,130)$ & $1.5$ & $(0.7,4.2)$ & - &  & - &  \\
$R_Q$ [$R_S$]& $0.046$ & $(0.001,0.27)$ & - &  & - & & - & \\
$R_F$ [$R_S$]& $> 0.02$ & &  $0.0013$ & $(0.0009,0.0020)$ & - &  & - & \\
p & -  & & - & & $2.88$ & $(2.82,2.94)$  & $2.4$ & $(2.1,3.1)$\\
 & & & & & & & &\\
$\chi^2$ / d.o.f. & 70.1 /74  & & 69.9 /74 & & 72.4 /77 & & 70.4/76 & \\
reduced $\chi^2$ & 0.95 & & 0.94 & & 0.94 & & 0.93 & \\ 
 & & & & & & & &\\
Violates $3\sigma$ MIR upper limit?& No &  & No &  & Yes &  & No &  \\ 
\enddata
\tablecomments{Summary of best fit parameters for different scenarios: synchrotron + submm IC (\texttt{icmodel}), synchrotron + NIR SSC (\texttt{sscmodel}), simple power law (\texttt{powerlaw}), power law with cooling break (\texttt{powerlawcool}). In each case a NIR $\nu L_\nu$ slope of $\beta = 0.4 \pm 0.2$ was enforced in order to add enough constraint to the parameters. We found that models which violated the NIR slope by $\gtrsim 2\sigma$ usually violated the 3$\sigma$ MIR limit also. Listed is also whether the model violates the MIR limit. Next to each value we provide the 90\% confidence interval for each parameter.}
\end{deluxetable}

\begin{figure*}
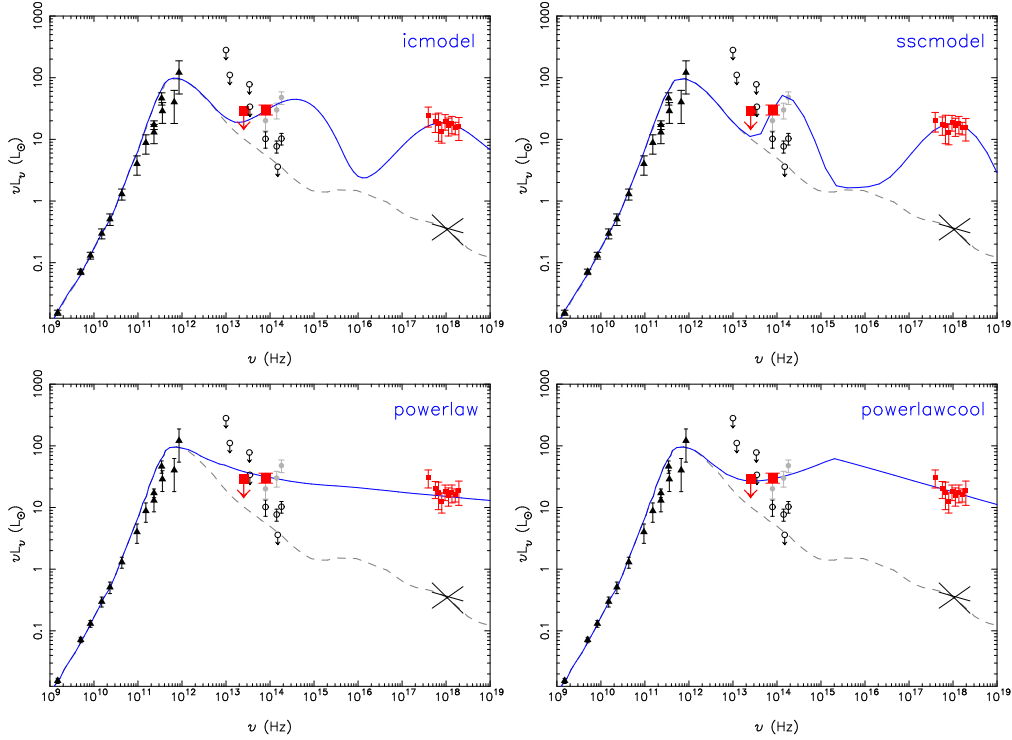

\begin{center}
\includegraphics[height=0.4\columnwidth, angle=-90]{f6a.eps}
\includegraphics[height=0.4\columnwidth, angle=-90]{f6b.eps}\\
\includegraphics[height=0.4\columnwidth, angle=-90]{f6c.eps}
\includegraphics[height=0.4\columnwidth, angle=-90]{f6d.eps}\\
\end{center}
\caption{The best fits of the four XSPEC models for the April 4 flare IR and X-ray data. The X-ray data points are the unfolded spectrum for the given model (blue solid line); only the PN unfolded spectra are shown, with the data points binned for plotting purposes. (i) \texttt{icmodel}: The solid blue line shows the best fit IC model for the fit constraints of the April 4 flare, which satisfies the MIR limit and the NIR spectral index. The dashed blue line shows the best fit model holding the magnetic field fixed at B=30 G. This model violates the NIR spectral index and comes close to violating the MIR upper limit. It does however allow a larger size for the quiescent region, $R_Q= 0.27$ $R_S$. This is however still far from the size/photon densities expected from size measurements of Sgr A*.
(ii) \texttt{sscmodel}: Best fit SSC model. In this case the magnetic field and density are extremely high. The source becomes self-absorbed in the NIR, and the spectrum shows strong curvature from L$'$ to H-band. (iii) \texttt{powerlaw}: Best fit power law model. This model violates both MIR limit and NIR spectral index. (iv) \texttt{powerlawcool}: A more feasible synchrotron model with a cooling break. This model corresponds to the steady state solution for a system with a constant injection of power law electrons where the energy loss of the electrons due to synchrotron emission is taken into account.}
\label{SED_allmodels}
\end{figure*}

Table \ref{Table_FittedParameters} lists the parameters of the fitted models. 
Figure \ref{SED_allmodels} shows the SED corresponding to the best fitting case for each model. In the next subsections we go through each model in detail.

\subsection{\textnormal{\texttt{icmodel}}: Flare caused by inverse comptonized submm bump photons}\label{ICsection}
A best fit model for the case of submm photons scattered by IR-emitting electrons is shown in Figure
\ref{SED_allmodels}. The model is a satisfactory fit to the data. The typical electron energies involved appear reasonable ($\gamma \sim 140$). However, the magnetic field strength of 210 G is high compared to that expected for the inner regions of the accretion flow ($10-30$ G), and the parameter $R_Q$ has a best fit value of $0.046$ $R_S$, which as we will argue is an unreasonably small size to contain the quiescent state of Sgr A* (stated another way, it corresponds to a much higher photon density of submm photons within the flaring region than can be expected).

To understand whether the small value of $R_Q$ is really `too small', we must be sure of how the parameter $R_Q$ should be interpreted. We note that $R_Q$ is in fact constrained through the quiescent photon energy density required to produce X-ray emission of the amplitude that we see in the X-ray flare. Since photon density is defined through $U_{ph} = L/cA$, with $A$ the surface area of the region emitting the luminosity $L$, we see that $R_Q$ can be interpreted as a constraint on the surface area of the (quiescent) emission region.

 Thus the true quiescent region's geometry must have a surface area equivalent to the surface area of a sphere of radius $R_Q$ in order to reproduce the required photon density. For example, a torus situated at the last stable orbit, $R_{\text{LSO}} = 3 R_S$ would have an equivalent surface area to our best fit $R_Q$ for a ring thickness of ${2 \times 10^{-4}}$ $R_S$: extremely thin.

New observations of the size of Sgr A* at 1.3 mm,  \citep{Doeleman2008_Nature} approaching the peak of the submm bump, suggest that the quiescent emission region may not be centred on the black hole. The reason for this is the fact that the measured size for Sgr A* at 1.3mm is smaller than the minimum apparent (gravitationally lensed) size allowed for an object very near a black hole. If the quiescent emission region is indeed offset from the position of the black hole then we no longer require an extremely thin ring containing most of the quiescent region electrons. However, this still does not solve the size issue: at the 90\% confidence level, the largest value of $R_Q$ that is compatible with the data, $R_Q = 0.27$ R$_S$ (0.54 R$_S$ in diameter) is still far below the measured FWHM size at 1.3 mm of $\approx 3.7$ R$_S$.

Another issue stems from the fact that our model does not take into account a stratified region \citep[i.e. the property that the observed size changes with wavelength,][]{Bower_etal_2004_Science, Shen_etal_2005_Nature} in the calculation of the photon density from $L_{\text{submm}}$ and $R_Q$. Within the true (non-homogeneous) quiescent source the local density of quiescent photons will change with position. It may be more realistic, rather than to input the photon density via the variable parameter $R_Q$, to input the known photon density of a known model for Sgr A* which satisfies all the observations including the size measurements. As a demonstration of how we can implement this we again take the model of \citet[][see Figure \ref{SED}]{Yuan_etal_2003} for which we have obtained tables of the quantity $n(\nu,R)$ at different radii $R$. This model has been shown to predict sizes at 3.5 and 7 mm consistent with those observed \citep{Yuan_etal_2006}.  

Implementing this model, the free parameter replacing $R_Q$ is the radial position of the flare from the the central black hole, $r_{pos}$. This position will determine the local photon energy density that is to be inverse Compton scattered by the flare electron population. When we incorporate the model's photon density into our inverse Compton code however, we find it difficult to find a reasonable fit for any $r_{pos}$. 

\begin{figure}
\includegraphics[width=\columnwidth]{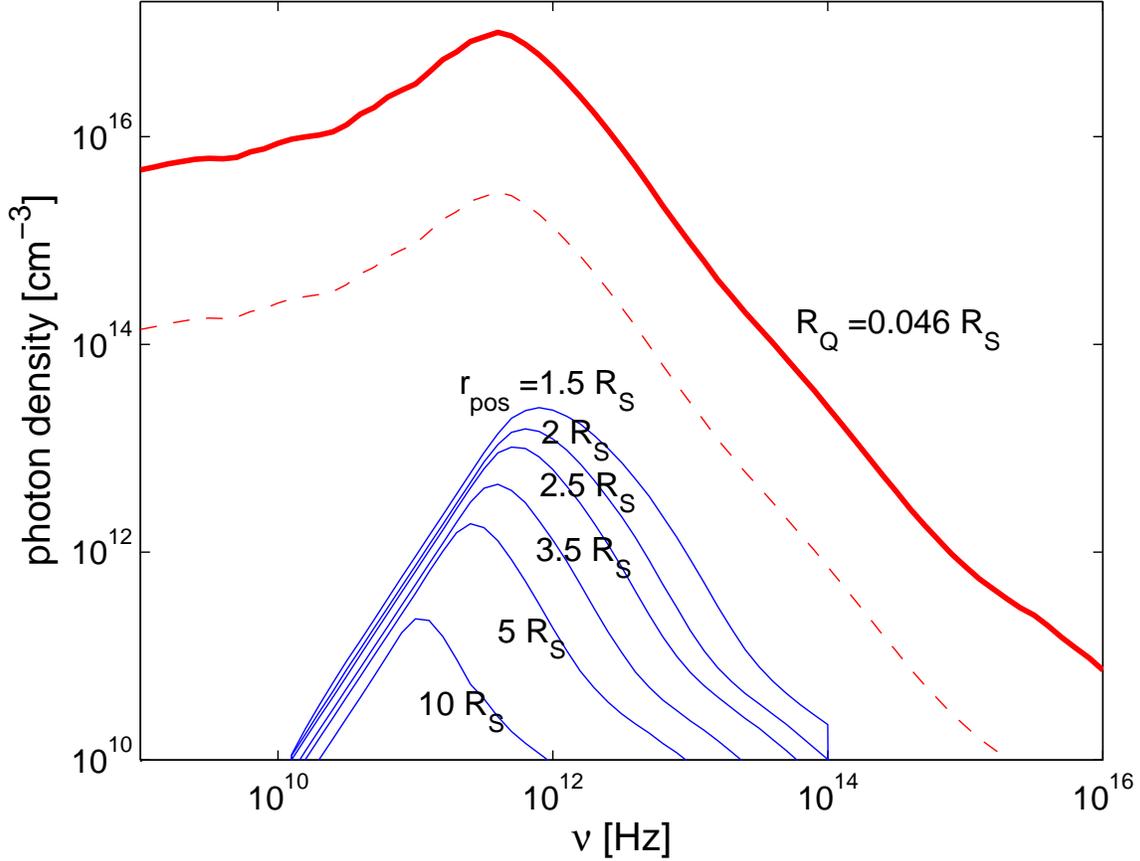}\\
\caption{Comparison of the photon density at each frequency in (i) the best fit IC model with $R_Q = 0.046 R_S$ (thick red solid line) and (ii) different radii between 1.5 $R_S$ and 10 $R_S$ in the \citet{Yuan_etal_2003} model (thin blue solid lines). The dashed red line shows the the photon density spectrum for $R_Q = 0.27$, at the $90\%$ confidence level for the parameter $R_Q$. The photon densities of \citet{Yuan_etal_2003} are in general 2 to 3 orders of magnitude too low, which shows why there is no well fitting IC model for the multiwavelength observations of the April 4 flare, given the photon densities of the \citet{Yuan_etal_2003} model.}
\label{photondensity}
\end{figure}

One can understand why this is through Figure \ref{photondensity}, which shows the energy density spectrum for (i) our simple `one-size' models, for the best fit $R_Q$ as well as a $\sim$ 3$\sigma$ upper limit (i.e. corresponding to the lowest energy density spectrum allowed to produce acceptable X-ray IC scattering) compared with (ii) the energy density spectra for various inner radii of the Yuan et al model. As we can see, the photon density in the model is just too low over the entire frequency range to reproduce a bright, soft, X-ray flare via inverse Compton scattering. 

Finally, although we were fitting an IC model to the flare, we must not forget about the fact that the electrons producing the NIR and IC emission must also be producing SSC emission. In fact, assuming the IC scenario to be the cause of the X-ray flare, we can put a lower limit on the size of the flare emission region $R_F$, by requiring the \emph{absence} of an SSC contribution. To estimate this, we slowly varied $R_F$ from its maximum value ($R_Q$) and observed at which point SSC emission began to overwhelm the IC X-ray emission\footnote{For the electron energies of our IC models, the SSC emission peak always occurred higher than X-ray frequencies, so the SSC spectrum had a hard $\nu L_\nu$ spectral index and is not a valid solution.}. We found that the flare region must be more than 0.02 $R_S$ in size. This would imply a density of $n_e \lesssim 8 \times 10^8 \text{cm}^{-3}$, which could be compatible with the kinds of densities ($n_e \approx 10^7 \text{cm}^{-3}$) expected for the inner regions of an accretion flow near Sgr A* \citep{Yuan_etal_2003}

\subsection{\textnormal{\texttt{sscmodel}}: Flare caused by inverse Comptonized NIR flare photons (SSC case)}

The critical frequency for synchrotron emission ($\nu_c$, the frequency at which a synchrotron-emitting electron emits most of its energy, and thus the energy at which the $\nu L_\nu$ spectrum turns over, see Section \ref{discussion_IC}) is linear in $B$ and quadratic in
$\gamma$. Thus to obtain an SSC peak below $10^{18}$ Hz the electrons must have very low energies of $\gamma \approx 10\text{-}15$. If the electrons were to have such low energies, it follows that a very large $B$ is needed to produce a synchrotron peak above NIR frequencies so that an increasing $\nu L_\nu$ spectral index is observed in the NIR.

Thus it makes sense that our best fit SSC model, shown in Figure
\ref{SED_allmodels}, corresponds to low electron energies and high magnetic field strength. The magnetic field strengths required are enormous, a factor of $\gtrsim 200$ greater than the typical magnetic fields of the quiescent state. At the same time, a dramatic decrease in the ratio of $\theta_E / B$
will have the effect of suppressing the SSC emission. 
Accordingly the density required from the small size, $n_e \approx 9 \times 10^{11}\text{cm}^{-3}$, is equivalent to a density enhancement on the order of $10^4$ above densities typical for the inner regions of the accretion flow and is also unrealistic. The high densities required even have the result that the synchrotron spectrum at IR/NIR wavelengths becomes self-absorbed. The self-absorption results in a very steep spectrum at IR wavelengths, and it shows significant curvature. Overall, due to the extreme physical conditions required to create the observed X-ray emission via SSC, we rule it out as the emission process behind the April 4 X-ray flare.

\subsection{\textnormal{\texttt{powerlaw}}: X-ray and IR flare from single power law synchrotron emission}

We have found that neither the IC nor SSC scenarios are entirely satisfactory as explanations for the simultaneous observations of April 4. Therefore, it is worthwhile to investigate other possible scenarios for the production of the X-ray flare. One such possibility is that both IR and X-ray flares are synchrotron emission.

A power law energy distribution of electrons
$$N(E)dE \propto E^{-p}dE $$
in the presence of a magnetic field, will create a synchrotron emission spectrum
$$\nu L_\nu \propto \nu^{(3-p)/2}.$$

While the \texttt{powerlaw} model gives a reasonable fit to the L$'$-band and X-ray data, it violates the MIR limit and gives a soft $\nu L_\nu$ spectral index in the NIR. This model is also unrealistic since for reasonable magnetic field strengths $B\approx 10-30$ G we expect electrons within the energy range of our power law distribution to have very short cooling timescales. Either very low ($B<0.1$G) magnetic fields are needed or very high ($B>60$G, together with a very flat spectrum of injected electrons $p\sim1.9$) to prevent a cooling break occurring between IR and X-ray wavelengths (i.e., this motivates our next model \texttt{powerlawcool}).  We can thus definitively rule out this scenario.

\subsection{\textnormal{\texttt{powerlawcool}}: X-ray and IR flare from power law synchrotron emission with cooling break}

It is well known for synchrotron emission sources to exhibit various breaks in their spectra due to cooling processes \citep[e.g.][]{Pacholczyk1970}.  One of the lowest frequency breaks likely to occur is due to synchrotron losses. The electrons responsible for the emission above this cooling break lose energy due to synchrotron cooling faster than they can typically escape (which they do on roughly the dynamical timescale). If the source of acceleration in the plasma occurs continuously (i.e. there is a continuous injection of electrons from the heating/acceleration process), a steady state solution exists where the spectrum follows the usual synchrotron spectral index of $\beta = (3-p)/2$ (with $p$ the particle index) at energies below a characteristic energy, the `cooling break', while above this energy the spectral index flattens to $\beta = (2-p)/2$, corresponding to a particle index of $p+1$ \citep{Yuan_etal_2003,Kardashev1962}. The position of the cooling break corresponds to the electron energy (or, in the emitted spectrum, the frequency) at which the cooling time is equal to the escape time. The escape time is uncertain but for our simple estimate we will use the dynamical time:
\begin{equation}\tau_{cool}  = 8 \left(\frac{B}{30\text{ G}} \right)^{-3/2}\left(\frac{\nu}{10^{14}\text{ Hz}} \right)^{-1/2}\text{ min}\label{Eq_coolingtimescale}\end{equation}
\begin{equation}\tau_{esc}\approx\tau_{dyn} = \sqrt{\frac{R^3}{2GM}} \approx 5 \text{ min}\end{equation}
with $R\approx 3.5$ $R_S$ the radial position within the accretion flow.
The cooling break therefore occurs at a frequency of
\begin{equation}\nu_{cool} = 2.56\left(\frac{B}{30\text{ G}}\right)^{-3}\times 10^{14} \text{ Hz}\label{Eq_coolingbreakfrequency}\end{equation}

This model has more freedom than the power law model, and the data provide less constraint on physical parameters than in the IC or SSC (because in those cases the IR and X-ray flares arise from different emission mechanisms). We can determine what magnetic field strengths are necessary for such a model since the magnetic field $B$ directly influences the position of the cooling break. 

Such a cooling break model fits the data well. The magnetic field strengths we find for this case are of the order of the magnitude of those expected for the inner regions around Sgr A*.

\section{Flare Evolution: Lightcurve Shape and Substructure}\label{section_flareevolution}

The SED modelling we presented in the last section only examined the mean properties of the flare emission. Our observations hold a great deal more valuable information in the time-dependent properties of the lightcurves.  It is interesting to examine what different emission scenarios imply for the evolution of simultaneous flare in the IR and X-ray bands. There are two outstanding features of the simultaneous lightcurves that need to be understood: the broadness of the NIR lightcurve in comparison to the X-ray lightcurve, and the substructures seen in the NIR lightcurve but not in the X-ray lightcurve.

The synchrotron and IC luminosities depend on $B$, $\theta_E$ (think $\gamma$)\footnote{We use $\theta_E$ and not $\gamma$ in our arguments because $\theta_E$ represents a characteristic energy of the entire population of electrons, while $\gamma$ more properly denotes the energy of each electron in the population.}, $N$ and the size of the quiescent region $R_Q$ as \begin{equation} L_{\text{synch}} \propto N\theta_E^2 B^2
\label{eq_syncLum}
\end{equation}
\begin{equation} L_{\text{IC}} \propto N\theta_E^2 R_Q^{-2} \label{eq_IClum}\end{equation}
The SSC luminosity depends not on $R_Q$ but on the size of the flaring region $R_F$ as \begin{equation} L_{\text{SSC}} \propto
N^2 \theta_E^4 B^2 R_F^{-2}. \label{eq_SSClum}\end{equation} 

The (changing) parameters governing the overall flare
evolution are probably $N(t)$ and the electron
temperature $\theta_E(t)$. For the case of submm photon IC, these are the only
parameters that can affect both the optically thin synchrotron
luminosity (not dependent on flare or quiescent region size) and the
IC scattered luminosity (not dependent on magnetic field).

For the submm IC case, both the synchrotron and inverse Compton luminosities depend on the same powers of $N$ and $\theta_E$. This means that if only $N$ or $\theta_E$ were to change throughout a flare, the X-ray lightcurve should follow the same functional form as the synchrotron lightcurve. That is, \emph{both lightcurves should have the same width, or duration}. 
We can take this either 
\begin{itemize}
 \item 
as further evidence against the submm photon IC scenario, or 
\item to imply that \emph{if} the X-ray flare is to be explained by the submm IC scenario, then the observed lightcurve widths can only be produced if some parameter aside from N and $\theta_E$ also varies throughout the flare.
\end{itemize}

If we consider the second possibility, then the magnetic field, $B$, is perhaps the most obvious choice for the varying parameter. The dependence of the lightcurves on $B$ means that for a change in $B$ to create a broader NIR lightcurve, the magnetic field must \emph{decrease} during the flare. The process must reverse itself towards the end of the flare: the magnetic field must increase again towards initial values. Such behavior could, for instance, occur if stored magnetic energy in a small region were released to accelerate electrons, as in a reconnection event. Remember however the high magnetic fields required ($B\sim200$ G) for the submm IC picture with \texttt{icmodel} in Section \ref{Section_SEDmodelling}, which makes it difficult to decrease the magnetic field during a flare unless we accept even higher values for the magnetic field before and after the flare. 

The only other possibility to explain the lightcurve durations in the submm IC scenario is that the photon density increases during the flare (i.e. effectively through the parameter $R_Q$). Although it is not realistic for the overall photon density of quiescent state photons to change much with time, the photon density experienced by the flare electrons could increase as it moves inwards within the accretion flow towards higher submm photon density. Again, to explain the second half of the flare, for this possibility the position of the flare within the accretion flow must move outwards towards lower photon density to increase the IR emission relative to X-ray emission as the flare declines.

As far as the substructure is concerned we can see from the same relations that a
variation in the magnetic field affects the synchrotron luminosity
but not the IC luminosity of submm-bump scattered photons. Within the IC picture then, it would be quite natural for the variations seen in the L$'$-band luminosity to be due to fluctuations in the magnetic field. The IC luminosity of submm bump photons, not dependent on $B$, would remain unaffected by such
fluctuations. If relativistic effects are also taken into account (Doppler boosting in particular) then the magnetic field fluctuations we are talking about are actually the fluctuations in the magnetic field of the \emph{observable} region at any given time.

  However, if the magnetic field were as high as that found in the SED modelling section for the IC model ($B\sim210$ G), then there would be a cooling break below IR frequencies and this picture could no longer work, since above the cooling break the lightcurve traces the pure rate of energy injection (no longer with any dependence on magnetic field). We are faced with a dilemma (in addition to the small size of the quiescent region which is required), where high magnetic fields are required to find an acceptable solution for the observed SED of the flare, but low magnetic fields are needed to keep the cooling break above IR frequencies, and we can not have both at once. 

If we now examine the SSC scenario, the SSC luminosity goes quadratically in the quantity $N\theta_E^2$, which means that a synchrotron lightcurve has a natural width that is a factor $\sqrt{2}$ times the width of its SSC emission (for example, if the lightcurves can be described by Gaussian profiles $f(t) \propto \exp(-k(t-t_0)^2/w^2)$. Thus a longer duration synchrotron lightcurve is expected in the synchrotron case.

The observations of substructure however are not naturally explained. Within the SSC scenario both the SSC luminosity and the synchrotron luminosity are
proportional to $B^2$, and fluctuations should thus induce variations of similar strength in both L$'$-band and X-ray lightcurves. As discussed in Section \ref{substructure}, substructure would be distinguishable in the X-ray lightcurve if it were of similar amplitude to that of the L$'$-band lightcurve. 

Finally, there is the synchrotron scenario with a cooling break. Understanding simultaneous light curves in this scenario is more sensitive to the time dependent evolution of the electron distribution itself and thus self-consistent time-dependent modeling is required. We can make at least some qualitative predictions for the light curves of this model:  as far as the difference in light curve widths is concerned, it seems it would be necessary for the cooling break to increase in frequency during the flare which would, as for the submm IC case, require a decrease in magnetic field during the flare (see Equation \ref{Eq_coolingbreakfrequency}). In contrast to the submm IC case, for this case the decrease in magnetic field could occur together with plausible values for the magnetic field. 

It also turns out that obtaining substructure in the IR light curve at the same time as producing a smooth X-ray light curve could be quite natural in the cooling break synchrotron model. Below the cooling break the emitted synchrotron spectrum is sensitive to variations in the magnetic field which is expected to be clumpy; while above the cooling break on the other hand the synchrotron emission traces rather the rate of energy injection alone. This could be expected to be rather smooth.

We think these aspects add very much to the plausibility of the synchrotron scenario as a viable mechanism for the production of the NIR/X-ray flare. 

\section{Discussion}

\subsection{Why inverse Compton scenarios don't work}\label{discussion_IC}

Here we give some analytical arguments that clarify which essential features of our multiwavelength observations lead us to exclude the inverse Compton processes as possible emission scenarios.

There are three equations \citep[see][]{Rybicki&Lightman1986} which essentially describe all important relationships between seed, synchrotron and scattered spectra involved in one inverse Compton scattering process. The first of these is the relationship that describes the shift in frequency of a seed photon upon encountering an electron of energy $\gamma$:
$$ \nu_{IC} = \gamma^2 \nu_{seed}$$ 
For the submm IC case, photons are scattered from the submm bump, at $\nu_{seed} \sim 10^{12}$ Hz. For the X-ray slope to be soft ($\nu_{IC}< 10^{18}$ Hz) this equation restricts the electron energies of the electrons involved in the flare:
$$\gamma \lesssim 1000.$$

Secondly, the frequency at which these electrons are themselves emitting synchrotron emission is dependent upon $\gamma$ and $B$ \footnote{Note that the formula we give here is for the case where electrons are spiralling exactly perpendicular to the magnetic field direction; i.e. the above equation actually contains a $\sin\theta$ term ($\theta$ the pitch angle of the electrons to the magnetic field) which is at maximum 1. If we took a smaller constant value (more realistic for an electron distribution with isotropically distributed pitch angles) then the results which follow would be even more restrictive.}:
\begin{equation} \label{nuc_equation}\nu_c = 4.2 \times 10^6 B \gamma^2.\end{equation}

Since we have already made a restriction on the electron energies, then if we require $\nu_c \gtrsim 10^{14}$ Hz (for a hard $\nu L_\nu$ spectral index in the IR), then we find that the magnetic field is also restricted:
$$B \gtrsim 25\text{ G}.$$
Thirdly, there is the equation relating the ratio of IC to synchrotron luminosity:
\begin{equation}
 \frac{L_{\text{compt}}}{L_{\text{synch}}} = \frac{U_{\text{ph,seed}}}{U_B}
 \label{ICequation}
\end{equation}
where $U_{\text{ph,seed}}$ is the energy density of seed photons,
$U_{\text{ph,seed}} \simeq L_{\text{ph,seed}}/c{A}$, and $A$ the surface area of the electron population producing the seed spectrum. 

In the submm case $L_{\text{ph,seed}}$ is $L_{\text{submm}}$, and the equation can be rewritten for $R_Q = \sqrt{A /4\pi}$ as 
\begin{equation}\label{RQ_equation_IC} R_Q \simeq 0.013 \left(\!\frac{L_{\text{F}}}{L_\odot}\!\right)^{\!\!\!1/2}\!\!\left(\!\frac{L_{\text{Q}}}{L_\odot}\!\right)^{\!\!\!1/2}\!\left(\!\frac{L_{\text{IC}}}{L_\odot}\!\right)^{\!\!\!-1/2}\! \left(\!\frac{B}{40\text{ G}}\!\right)^{\!\!\!-1} R_S \end{equation}

With this we obtain an uncomfortably low constraint on
the size of (the most luminous part of) the quiescent region of $R_Q \lesssim
0.1 R_S$, similar to the small sizes we found were required in the SED model fitting of Section \ref{Section_SEDmodelling}.

We can repeat the above series of arguments for the SSC case to obtain \begin{eqnarray*}
\gamma &\lesssim& 100\\
B &\gtrsim& 2400 \text{ G}\\
R_F &\leq& 0.002 \text{ R}_S
\end{eqnarray*}

A constraint on the size of the flare emitting region itself also enables a constraint on the density of the flare emission region: $$n_e \simeq 2.4 \times 10^9 \left(\!\frac{L_{\text{F}}}{L_\odot}\!\right)^{\!\!\!-2}\!\left(\!\frac{L_{\text{SSC}}}{L_\odot}\!\right)^{\!\!\!3/2}\!\left(\!\frac{B}{40\text{ G}}\!\right)\!\left(\!\frac{\gamma}{100}\!\right)^{\!\!\!-2} \text{cm}^{-3} $$
and we obtain that $n_e \gtrsim 10^{10}$ cm$^{-3}$. As with the density we found in the SED modelling of Section \ref{Section_SEDmodelling}, this is a very high density, several orders of magnitude higher than the density inferred for the inner regions of the accretion flow around Sgr A*\citep[$\sim 10^7$ cm$^{-3}$:][]{Yuan_etal_2003}.

As for the submm IC case, the driver for the high magnetic fields and densities in the SSC case is the restriction on the electron energies required by the soft X-ray slope, which in combination with a positive IR slope forces the magnetic field to very high values. Adding to that the high luminosity ratio, $L_X/L_{IR}$, the size of the flare region is driven to very small values which in turn forces the density to very high values. 

These arguments show that for both cases there are three main properties which in combination are driving (i) in the submm IC case, the magnetic field to higher values than $B\approx30$ G and the IC region to small sizes, and (ii) in the SSC case, to extremely high magnetic fields and electron densities. These are:
\begin{enumerate}
 \item the soft X-ray spectral index
 \item the hard MIR-IR spectral index 
 \item the high luminosity ratio, $L_X/L_{IR}$.
\end{enumerate}

\subsection{Comparison with past multiwavelength studies}

Past multiwavelength observations of Sgr A* have favored models where the X-ray emission is due to inverse Compton scattering processes, with various possible combinations of seed photons and seed electrons from those producing the quiescent (submm) and flaring (IR) states. The obvious question is: why do we not find the same?

\citet{Eckart_etal_2004} modelled both IR and X-ray flares as SSC emission. However, that the flares at IR wavelengths are SSC emission is ruled out since the polarization of the IR flares points to a synchrotron origin.

\citet{Eckart_etal2006_Multiwavelength} modelled the X-ray flare as SSC emission via the prescription of \citet{Marscher1983}, and in their model the X-ray emission had a hard $\nu L_\nu$ index of $\beta=0.4$ to match their X-ray observations. A hard $\nu L_\nu$ index in the X-ray will provide no constraint on the energies of the electrons producing the IR flare, which explains why these studies found this model to be acceptable with reasonable sizes for 
the quiescent state of Sgr A*. The X-ray flares of their study were however quite weak and the photon index may not have been well constrained \citep[][ Mascetti et. al. 2008 (submitted)]{Porquet_etal_2008}.

A similar story applies to other investigations of the X-ray flares with IC and SSC processes. For example, in \citet{LiuMeliaPetrosian_etal_2006} and \citet{Marrone2008} the X-ray emission was modelled with SSC, but again this was for a hard spectral index in $\nu L_\nu$. For these cases as for \citet{Eckart_etal2006_Multiwavelength} there was no restriction on the electron energies since the X-ray flare was taken to be hard, and accordingly it was possible to find a model with reasonable physical parameters. 

\citet{YusefZadeh_etal_2006} suggested two scenarios where the X-ray emission could be due either to (i) submm photons upscattered by electrons producing IR emission, the scenario we considered in Section \ref{ICsection}, or
(ii) IR (flare) photons upscattered by the electrons producing the quiescent state of Sgr A*. 

The first case is the case we explored in Section \ref{ICsection}. In \citet{YusefZadeh_etal_2006} however, the treatment involves a power law distribution of electrons and the differential flux is calculated for the corresponding power law section of the inverse Compton scattered spectrum only. Since the X-ray spectral index in this model was considered to be hard in $\nu L_\nu$ ($\beta=0.4$), we have the same situation as with the previous cases; there is no restriction on the electron energies producing the IR flare. Therefore, with electrons up to $\gamma \sim 6000$, it was possible to find a model that worked using a reasonable quiescent region size ($R_Q \approx 10$ $R_S$).

The second case is a scenario we did not consider in our modelling. In fact it can be shown that the inverse Compton luminosity via this process (IR photons scattered by submm-emitting electrons) can never exceed the luminosity of the IC case we considered previously (submm photons scattered by IR photons).

For the case of IR photons scattered by submm-emitting electrons, the luminosity is (here using $\nu$ to denote a photon)
$$L_{IC} (\text{IR }\nu\text{, submm e}^-) = \frac{2 R_F L_{submm}L_{IR}}{R_Q^3 c B^2}  $$
where we have made use of the fact that only some proportion of the quiescent electrons (those within the flaring region where the photon density of IR photons is highest) are available to inverse Compton scatter photons:
$$L_{synch} = \frac{R_F^3}{R_Q^3} L_{submm}. $$
For the submm IC case that we already considered we have
$$L_{IC}(\text{submm }\nu\text{, IR e}^-) = \frac{2 L_{submm}L_{IR}}{R_Q^2 c B^2}  $$
and the ratio of the two is 

$$\frac{L_{IC} (\text{IR }\nu\text{, submm e}^-)}{L_{IC}(\text{submm }\nu\text{, IR e}^-)} = \frac{R_F}{R_Q}$$

Thus the X-ray luminosity provided by the IR seed photon case is always going to be, at best, comparable to the X-ray luminosity produced in the submm seed photon case and will never dominate the emission. In Section \ref{ICsection} we found it was difficult to find a solution that did not involve an unrealistically small size for the quiescent region of Sgr A*. The contribution of IC emission through the scattering of IR seed photons by submm-emitting electrons is at best comparable to this emission and can not satisfy the observations either.

We also compare expectations for the inverse Compton scattered vs. synchrotron lightcurves for this scenario. The different widths of the lightcurves is as difficult to understand as in the case of inverse Compton scattered submm bump photons; again, the only way out may be for the magnetic field to decrease during the flare, and to be restored at the end of the flare. Additionally, if the IR flare provides the seed photons for the X-ray flare, then we should expect to see substructures in the X-ray lightcurve of the same order as and simultaneous with those in the IR lightcurve. Fluctuations in the magnetic field could not help this scenario because the IR flare is directly providing the seed photons for the X-ray flare in this case. Therefore we conclude that the time-resolved features of the lightcurves do not support this emission scenario either.

Overall, we have covered all reasonable conceivable inverse Compton scenarios for the origin of the X-ray flare simultaneous with our IR flare and have concluded that none of these inverse Compton scattering scenarios are viable.

Of past investigations, the most similar synchrotron model to those we present in this paper is the one-component synchrotron model presented by \citet[][see Figure 3]{Yuan_etal_2004}. To our knowledge, this is the only previous work to suggest that both IR and X-ray flares are produced by synchrotron emission. Interestingly, in that paper, a one-component model for the electron distribution is ruled out on the basis that the X-ray spectrum is too soft ($\Gamma \sim 2$) to be consistent with the X-ray flares observed by Chandra.

\subsection{The electron energy distribution}
In Section \ref{Section_SEDmodelling} where we investigated inverse Compton scenarios for the simultaneous IR/X-ray flare (\texttt{icmodel} and  \texttt{sscmodel}) we assumed a thermal distribution for the energy distribution of the relativistic electrons. The reason for this choice was for ease of calculation since the synchrotron emission of a thermal distribution of electrons is described by well known formulas (see Appendix). The thermal distribution is also an expected result of turbulent heating and radiative cooling processes and has been used elsewhere in models for the flares of Sgr A* \citep{LiuMeliaPetrosian_etal_2006}. Nevertheless, one might worry that through our assumption of a thermal distribution of electron energies we have limited ourselves to a special case and that there may exist other distributions of electron energies that allow the inverse Compton scenario to be an explanation for the April 4 IR/X-ray flare after all.  Although we do not expect our results to be particular to the thermal distribution given that the arguments we developed in Section \ref{discussion_IC} were not specific to any particular arrangement of the electrons, we would like to here corroborate this expectation through modeling. 

\begin{figure}
\begin{center}
\includegraphics[width=0.4\columnwidth, angle=-90]{f8a.eps}
\includegraphics[height=0.47\columnwidth]{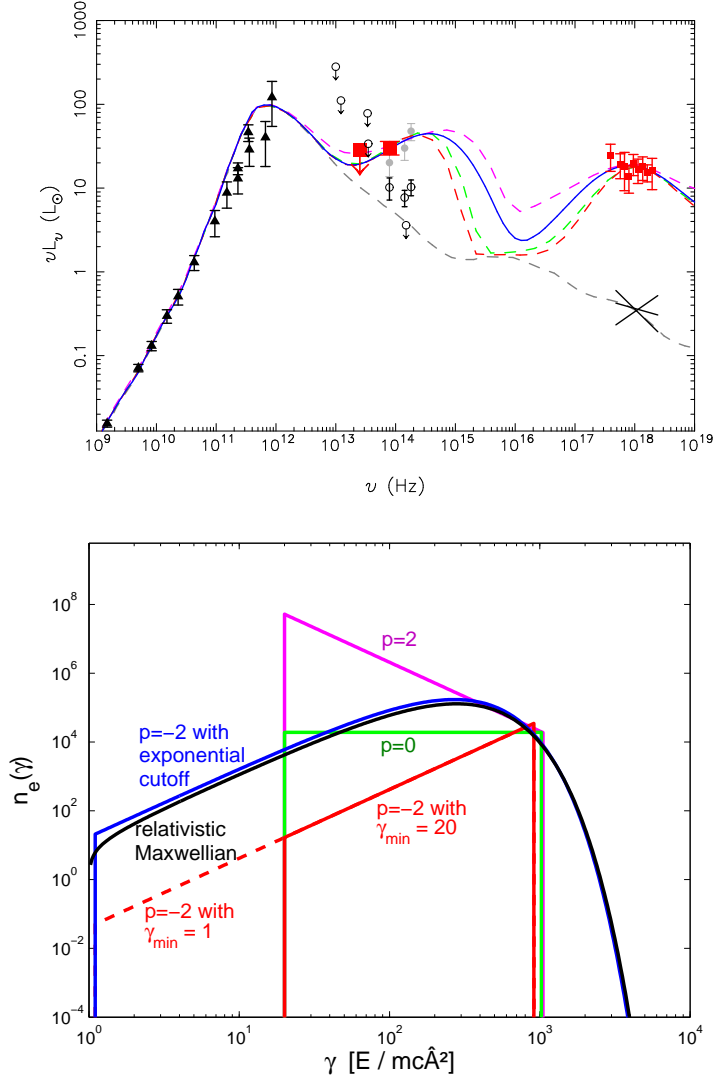}
\end{center}
\caption{\emph{Upper panel:} Best fit SEDs where the NIR synchrotron emission is produced by a power law distribution of electrons with index $p = 2,0,-2$ (magenta, green and red dashed lines). The best fit model for a thermal distribution of electrons is shown in solid blue. All models appear relatively similar. Note that the (model-dependent) X-ray data is only shown for the thermal distribution of electrons case, but it should be similar for the other models. \emph{Lower panel:} The electron energy distributions of best fit submm IC models each assuming a different underlying form for the energy distribution. We compare power law models ($n(\gamma) \propto \gamma^{-p}$) of index p = 2, 0 and -2 (magenta, green and red respectively). We also demonstrate that in a power law model of index -2 with an exponential cutoff (blue) we found the same energy distribution as in the thermal distribution case (in black; note that we used analytical equations to calculate the synchrotron emission for the thermal case in \texttt{icmodel} and \texttt{sscmodel} of Section \ref{Section_SEDmodelling}). Although the electron distributions all look very different, they all cross at an energy $\gamma\approx1000$. }
\label{energydists}
\end{figure}

To investigate the senstivity of our results to the form of the electron distribution we investigated power law models for the NIR synchrotron emission with the inclusion of inverse Compton scattering. We take power law models of the form
$n(\gamma)\propto \gamma^{-p}$ between energies of $\gamma_{min}$ and $\gamma_{max}$ and calculate the synchrotron and inverse Compton emission by integrating the synchrotron spectrum of a single electron of energy $\gamma$ over all electrons in the distribution. We fixed the parameter $\gamma_{min} = 20$, assuming that the electrons are accelerated out of the population of electrons ($\gamma \approx 10-20$) that create the quiescent state synchrotron emission, but we also found that the results are not at all sensitive to the value of $\gamma_{min}$ as we show in Figure \ref{energydists} where we show also a model with $\gamma_{min}=1$. As we did for our previous set of models, we fit the calculated SEDs to the observed data using XSPEC, naming this model \texttt{powerlawicmodel}. 

We tested three models, with $p=2$, $p=0$ and $p=-2$, corresponding to falling, flat and rising electron distributions respectively. In Figure \ref{energydists} we show the emitted SED and the electron distributions that correspond to the best fit in each case. The distribution of electron energies is irrelevant to finding a good fit to the observed data, and the SEDs for the best fits for all models appear similar. Amongst the best fit models the electron distributions show a lot of variety. For a power law of electrons with index $p=-2$ to which we also add an exponential cutoff, we find that the electron distribution of the best fit approaches the thermal distribution corresponding to the best fit for \texttt{icmodel}, as we would expect. 

Interestingly, for this variety of models, the electron distributions all meet at a particular energy, $\gamma\approx1000$. It appears that this energy is important (\emph{even, the only important energy}) for the submm IC scenario, and we note that it corresponds exactly to the upper limit on electron energy that we found in Section \ref{discussion_IC}, for the submm IC case.

\begin{deluxetable}{lll} 
\tabletypesize{\scriptsize}
\tablecolumns{3}
\tablewidth{0pc}
\tablecaption{Models: Fit Parameters\label{Table_powerlawicmodel}}
\tablehead{
\multicolumn{3}{c}{Fit to Mean Fluxes \& NIR spectral index} \\Parameter & \multicolumn{2}{c}{\texttt{powerlawicmodel}}} 
\startdata
N$_H$ [$\times 10^{22}\ \textnormal{cm}^{-2}$]& $11.8$ & $(10.0,14.3)$ \\
B [Gauss]& $330$ & $(22,2100)$ \\
$\gamma_{max}$ [\ ]& $1100$ & $(280,1600)$\\
$N_e$ [$\times 10^{40}$ electrons]& $20$ & ($1.7,1900)$\\
$R_Q$ [$R_S$]& $0.028$ & $(0.00003,0.32)$\\
$R_F$ [$R_S$]& $> 0.02$ & \\
p & 2.0  & \\
 & & \\
$\chi^2$ / d.o.f. & 70.3 /75  & \\
reduced $\chi^2$ & 0.94 & \\ 
 & & \\
Violates $3\sigma$ MIR upper limit?& No &  \\ 
\enddata
\tablecomments{Summary of best fit parameters for \texttt{powerlawicmodel} fixing $p=2$. The constraints on the parameters are very similar to those found for the $\texttt{icmodel}$ of Section \ref{Section_SEDmodelling} and appear to obey the constraints of Section \ref{discussion_IC}.}
\end{deluxetable}

In Table \ref{Table_powerlawicmodel} we give parameter values for the power law model with $p=2$. We find that in general the parameters, in particular the confidence intervals, are similar to those in Table \ref{Table_FittedParameters} for \texttt{icmodel}. Of particular interest is to see whether a larger value of $R_Q$ can be accommodated to revalidate the inverse Compton scenario. The upper limit (90\% confidence) for this parameter however of $R_Q < 0.32$ however, is still very small and does not allow for the this scenario to be compatible with the size measurements of Sgr A*. It appears that our finding for the submm IC scenario, where we find the implied size for the submm-emitting region must be much smaller than the observed size, is robust and is not sensitive to the form of the electron distribution.

Beyond assuming a particular form for the electron distribution, we should keep in mind that in reality the electron energy distribution is an evolving function that depends on the details of injection/escape and electron cooling. A truly self-consistent approach would allow the electron distribution function to be determined from the parameters themselves (such as, e.g. the magnetic field, which determines how long the electrons take to cool and thus has an important effect on the shape of the electron spectrum). The synchrotron spectrum would then be calculated directly from the time-dependent energy distribution function.

As an example of why the electron distribution should be self-consistently calculated from the model parameters, we note that both of our inverse Compton models require high magnetic fields. Even for the submm IC model with the lower of the two magnetic fields, the cooling timescale (Equation \ref{Eq_coolingtimescale}) is shorter than the dynamical timescale already at IR frequencies.  The electron distribution in the case of such high magnetic fields would then be already dominated by cooling at L$'$-band wavelengths. In essence, a further requirement on our models is that we hope to find a solution with $B~\lesssim 30$~G in order to keep the cooling break above the NIR band. However, as we have found in our analysis, there are certainly no solutions for the SSC model with such low magnetic field, and the submm IC only accommodates such a low magnetic field at the edge of its $90\%$ confidence region. 

\subsection{The jet model}

The model spectrum we used to calculate the photon density of submm photons in $\texttt{icmodel}$ was that of \citet{Yuan_etal_2003}. However, this is not the only model that describes the quiescent spectrum of Sgr A*; the quiescent spectrum is also well described by a jet model \citep{Falcke&Markoff_2000}.

\begin{figure}
\begin{center}
\includegraphics[width=0.4\columnwidth, angle=-90]{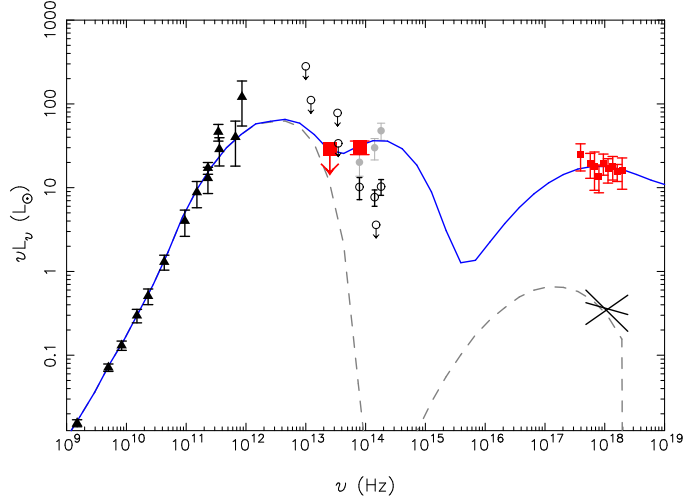}
\end{center}
\caption{A submm inverse Compton model for the X-ray flare using a different model \citep[the jet model, ][]{Falcke&Markoff_2000} as template for the submm photon spectrum. The dashed gray line shows the jet model which is used as the source of seed photons. The solid blue line shows the best fit XSPEC model. }
\label{jetmodel}
\end{figure}

To test whether our conclusions were robust with respect to the assumption of a quiescent model spectrum we created a new XSPEC model in which the submm spectrum to be upscattered was that of \citet{Falcke&Markoff_2000}. Figure \ref{jetmodel} shows the best fit model for this case, and the best fit parameters are very similar to those of $\texttt{icmodel}$, with somewhat higher magnetic field ($B \approx 300$ and lower electron temperature $\theta_E \approx 80$). The size of the quiescent region, $R_Q$ is still very small ($R_Q \approx 0.02$) for the best fit model. Thus we can conclude that our findings are also not sensitive to the particular model we used to model the submm photon density.

\subsection{Substructure in the context of an orbiting hot spot model}
It is worth making a closing comment on our results in the context of the orbiting blob model \citep{Genzel_etal2003_FirstNIRFlare, BroderickLoeb_2005, Meyer_etal_2006_hotspotmodel, Trippe_etal2007_PolarizedFlare}, where the substructures observed commonly in the IR/NIR lightcurves are postulated to be due to relativistic beaming on the approaching side of the hot spot's orbit about the SMBH.

Due to the fact that the X-ray lightcurve is so smooth, and the substructure only shows up in the NIR lightcurve, then naively, the hot spot model does not seem to be compatible with the observed lightcurves; if relativistic beaming is occurring, it is not obvious how such structures could \emph{not} also show up in the X-ray lightcurve.

We think this aspect of the multiwavelength emission from Sgr A* certainly deserves consideration in models describing the flare emission from Sgr A* via a hot spot model. The different widths of the lightcurves are an additional important clue in this regard. Conversely, it is also obvious that given its likely proximity to the SMBH of $\approx 4\times10^6$ $M_\odot$ relativistic effects can have a considerable influence on the observed emission from Sgr A* and should be taken into account in models that try to explain multiwavelength properties of flares from Sgr A*. 

\section{Conclusions}
We have presented the results of a simultaneous multiwavelength
campaign at L$'$-band, X-ray and MIR wavelengths carried out on April 4 2008.
We summarize the main observational results as follows:
\begin{itemize}
 \item The L$'$-band and X-ray flares were simultaneous to within 3 minutes.
 \item The L$'$-band flare is much broader overall than the X-ray lightcurve.
 \item The L$'$-band flare showed significant substructure with a timescale of $\sim$20 minutes, while the X-ray flare showed no corresponding significant substructure.
 \item The $\nu L_\nu$ spectrum increases between 11.88 $\mu$m and 3.8 $\mu$m.
 \item The X-ray flare was very bright and soft in $\nu L_\nu$ \citep[$\beta = -0.3 \pm 0.3$; ][]{Porquet_etal_2008}.
\item The emission region must be small, $< 1.5 R_S$.
\end{itemize}

We have drawn conclusions about the emission mechanism behind the flares which can be traced essentially to the hard MIR-IR $\nu L_\nu$ spectral index, the soft X-ray $\nu L_\nu$ spectral index, and the high ratio of X-ray to IR luminosity. Our conclusions are:
\begin{itemize}
\item We strongly disfavor the SSC case due to the extremely high magnetic fields and electron densities required to reproduce the observed data. Both quantities are several orders of magnitude larger than the values expected for the inner regions of the accretion flow about Sgr A*.
\item We disfavor the submm IC case due to the high submm photon densities required to produce the observed X-ray emission, which imply a quiescent region size much smaller than implied by the size measurements. The IC case is also only marginally compatible with physically plausible magnetic fields of $10-30$ G. 
\item We also disfavor a different IC case (where the X-ray flare is due to IR seed photons upscattered by quiescent state submm-emitting electrons) since the inverse Compton scattered luminosity in this case is always dominated by the inverse Compton scattered luminosity of submm photons scattered by IR-emitting electrons.
\item We favor a synchrotron scenario where the emitted spectrum flattens towards X-ray energies due to a cooling break. This scenario can be achieved with physically plausible magnetic field strengths and holds promise to explain more detailed structures of the lightcurves.
\\
\item For both the synchrotron and IC cases, the relatively shorter duration of the X-ray flare and perhaps the substructure of the IR emission is most plausibly understood as a result of a transient decrease of the magnetic field in the region of the flare. Such a decrease may point toward flares being triggered by conversion into electron heating of stored magnetic energy, such as a magnetic reconnection event.
\end{itemize}

\acknowledgements{\small{ \emph{Acknowledgements.} The XMM-Newton project is an ESA Science Mission with instruments and contributions directly funded by ESA Member State and the USA (NASA). We also received the support of PHASE, the high angular resolution partnership between ONERA, Observatoire de Paris, CNRS and University Paris Diderot Paris.}}

\bibliography{ms}

\begin{appendix}
\section{\texttt{icmodel} and \texttt{sscmodel}}
Both \texttt{icmodel} and \texttt{sscmodel} are subsets of a single model with five parameters $B$, $N$, $\theta_E$, $R_F$ and $R_Q$.

For this model, we assume a spherical homogeneous emission region (containing the transiently heated/accelerated electron (flare) population emitting in the IR) of size $R_F$ and of electron density $n_e=N/(4\pi R_F^3)$. The region contains a homogeneous magnetic field of strength $B$. We choose to model the electron distribution with the thermal electron distribution (i.e., a relativistic Maxwellian)
$$n(\gamma) = \frac{n_e \gamma^2 \sqrt{1-1/\gamma^2}}{\theta_E K_2(1/\theta_E)} \exp(-\gamma/\theta_E)$$
with $\theta_E = kT_e / m_ec^2$ the dimensionless electron
temperature, and $K_2(x)$ a modified Bessel function of the second kind. The thermal distribution is a good example of an electron distribution of a characteristic energy (i.e. $\gamma\sim\theta_E$), and is an expected result of turbulent heating processes \citep{LiuMeliaPetrosian_etal_2006}.\\
The emission coefficient is approximated by \citep{Mahadevan_etal_1996}
$$j_{\nu,\text{th}} = \frac{n_e e^2}{\sqrt{3}cK_2(1/\theta_E)}\nu
M(\frac{2\nu}{3\nu_b\theta_e^2}), $$in units of $\text{erg s}^{-1}\text{cm}^{-3} \text{Hz}^{-1} \text{ster}^{-1}$, and with
\begin{eqnarray*}
M(x) = 4.0505 a x^{-1/6}\exp(-1.8896 x^{1/3}) ( 1+0.40 bx^{-1/4}+0.5316cx^{-1/2}).\end{eqnarray*}
The absorption coefficient is
$$\alpha_{\nu,th} = j_{\nu,th} / {B_\nu(T_e)} = j_{\nu,th}~\frac{c^2}{2h\nu^3}~\left(\exp\left({h\nu} / {kT_e}\right)-1\right) $$
and the resultant synchrotron spectrum is computed for our spherical flare region (including optical depth effects) by the equation of radiative transfer
$$L_{\nu,\text{S}} = 4\pi \int_0^{R_F} \frac{j_\nu}{\alpha_\nu}\left(1-\exp(-\alpha_\nu\sqrt{{R_F}^2-r^2}\right)4\pi r dr $$
which for small optical depth (an optically thin flare) simplifies to
$$L_{\nu,S} = 4\pi\left(\frac{4\pi}{3} R_F^3 \right)j_\nu.$$
We compute the Inverse Compton scattered luminosity in our model
through \citep{Bluementhal&Gould_1970}
$$L_{\nu,\text{IC}} = \frac{4\pi}{3}R_F^3 (h\nu)^2\! \int_\gamma\! n(\gamma)\!\int_\epsilon(dN_{\gamma,\epsilon}/ dt d\epsilon_1 d\epsilon) d\epsilon d\gamma$$
where $\gamma$, $\epsilon$ and $\epsilon_1 = h\nu$ are the electron energy, initial and scattered photon energies respectively, and the quantity
\[  \begin{array}{r l}
    &dN_{\gamma,\epsilon}/ dt d\epsilon_1 d\epsilon = 3\sigma_T c~n_{ph}(\epsilon) / {4\gamma^2\epsilon}\left[2q\ln{q} + (1+2q)(1-q)+\frac{(\Gamma_e q)^2(1-q)}{2(1+\Gamma_e q)}\right]
  \end{array}\]
where $\Gamma_e = 4\epsilon\gamma / mc^2$ is the Compton factor, $q = \epsilon_1 / \Gamma_e(\gamma mc^2 - \epsilon_1)$.
The photon density $n_{ph}(\epsilon)$, of flare state photons is
determined from the model flare's luminosity by $n_{ph}(\epsilon) =
n_{ph}(\nu)/h = L_\nu/(4\pi h^2\nu c R_F^2)$. Instead of modelling the quiescent state ourselves, we use the
luminosity given by the quiescent model of \citet{Yuan_etal_2003}, which reproduces the observations quite well. Using this quiescent model as input
spectrum $L_{\nu,Q}$ then, the photon density of quiescent state
photons is $n_{ph}(\epsilon) = L_{\nu,Q}/(4/3\pi h^2\nu c {R_Q}^2)$. 

We implement \texttt{icmodel} and \texttt{sscmodel} via the above prescription, with only one difference concerning the parameters $R_Q$ and $R_F$. For \texttt{icmodel}, $R_F$ is not an input parameter, and is merely set to $R_F = R_Q$ (this simply ensures that a minimum of SSC emission is produced, so that a solution where the X-ray emission is due to inverse Compton scattering of submm photons may be found). Similarly for \texttt{sscmodel}, $R_Q$ is not an input parameter, but is set to an arbitrarily high value (in this case to ensure that very little inverse-Comptonized submm emission is produced and a solution is found where the X-ray emission is due to SSC of IR/NIR photons).

\section{\texttt{powerlawcool}}
This model can be written as
\[ \nu L_\nu \propto \left\lbrace
  \begin{array}{l l}
     \nu^{(3-p)/2} & \hspace{1cm} {\nu_{\text{min}}< \nu< \nu_c},\\
     \nu^{(2-p)/2} & \hspace{1cm} \nu_c < \gamma < \nu_{\text{max}}.
  \end{array}
\right. \]

The cooling break, $\nu_c$ (Equation \ref{Eq_coolingbreakfrequency}) occurs at
$\nu_c =  2.56\left(B/30\text{ G}\right)^{-3}\times 10^{14} \text{ Hz}$, and the index $p$ corresponds to the particle index in the underlying electron distribution:
\[ n(\gamma) \propto \left\lbrace
  \begin{array}{l l}
     \gamma^{-p} & \hspace{1cm} {\gamma_{\text{min}}< \gamma< \gamma_c},\\
     \gamma^{-(p+1)} & \hspace{1cm} \gamma_c < \gamma < \gamma_{\text{max}}.
  \end{array}
\right. \]
\end{appendix}

\end{document}